\documentclass[aps,twocolumn,superscriptaddress,nofootinbib,a4paper,longbibliography]{revtex4-2}
\pdfoutput=1

\usepackage{times}
\usepackage{epsfig}
\usepackage{amsfonts}
\usepackage{amsmath}
\usepackage{amsthm}
\usepackage{amssymb}					
\usepackage{amsthm}
\usepackage{dsfont}
\usepackage{bm}
\usepackage{mathtools}
\usepackage{color}
\usepackage{multirow}
\usepackage[normalem]{ulem}
\newcommand{\stkout}[1]{\ifmmode\text{\sout{\ensuremath{#1}}}\else\sout{#1}\fi}
\usepackage{latexsym}
\usepackage{mathrsfs}
\usepackage{natbib}
\usepackage{verbatim}
\usepackage[T1]{fontenc}
\usepackage{float}
\usepackage{graphicx}
\usepackage{xcolor}
\usepackage{physics}
\usepackage{soul}

\newtheorem{theorem}{Theorem}

\newtheorem{lemma}[theorem]{Lemma}

\newtheorem{result}[theorem]{Result}

\theoremstyle{definition}

\newcommand{\bracket}[3]{\langle#1|#2|#3\rangle}
\newcommand{\expect}[1]{\langle#1\rangle}

\usepackage[colorlinks=true,linkcolor=blue,citecolor=magenta,urlcolor=blue]{hyperref}

\begin{document}


\title{Quantum steering with imprecise measurements}

\author{Armin Tavakoli}
\affiliation{Physics Department, Lund University, Box 118, 22100 Lund, Sweden}

\begin{abstract}
We study quantum steering experiments without assuming that the trusted party can perfectly control their measurement device. Instead, we introduce a scenario in which these measurements are subject to small imprecision. We show that small measurement imprecision can have a large detrimental influence in terms of false positives for steering inequalities, and that this effect can become even more relevant for high-dimensional systems. We then introduce a method for taking generic measurement imprecision into account in tests of bipartite steering inequalities. The revised steering bounds returned by this method are analytical, easily computable, and are even optimal for well-known families of arbitrary-dimensional steering tests. Furthermore, it applies equally well to generalised quantum steering scenarios, where the shared quantum state does not need to be separable, but is instead limited by some other entanglement property.
\end{abstract}

\date{\today}

\maketitle

\textit{Introduction.---} Entanglement is both a fundamental quantum phenomenon and a paradigmatic resource for quantum information processing. Therefore, much research has been invested into detecting entanglement properties of initially uncharacterised states. Several routes have been developed to study this problem. In the most common approach, the experimenter performs a number of well-chosen local measurements and  analyses the data via an entanglement witness \cite{Guhne2009}. While this method in principle can  detect every entangled state, it is based on the idealisation that the experimenter can \textit{exactly} perform the prescribed measurements. A diametrically different approach sees the experimenter performing measurements that lead to the violation of a Bell inequality \cite{Brunner2014, Networks}. This way,  entanglement can be detected without requiring any characterisation of the quantum measurements. However, while the assumptions are minimal, they come with severe fundamental and practical limitations on the detectable states \cite{Augusiak2014}.

An intermediate scenario, known as quantum steering, has received much attention \cite{Uola2020}. In the steering scenario, the measurements of one party are uncharacterised  whereas those of the other party are known exactly. In principle, this makes steering experiments (see e.g.~\cite{Saunders2010, Wittmann2012, Händchen2012, Armstrong2015, Sun2016}) more demanding than entanglement witnessing but far less demanding than Bell nonlocality.  In contrast to what is known for nonlocality, steering has favourable noise and loss properties, which improve significantly for large Hilbert space dimensions. Therefore, recent years have seen much interest in high-dimensional steering experiments \cite{Zeng2018, Wang2018, Designolle2021, Srivastav2022, Qu2022, Qu22b, Huang2021}, which can now range far into double-digit dimensions.

However, alike standard entanglement witnesses, the steering scenario suffers from the need for one party to perform exactly known measurements. Such flawless control  is an assumption to which experiments can only aspire. Furthermore, on many relevant physical platforms, precise control of measurements is considerably  more challenging when scaling the dimension. In this work, we introduce a framework for steering without assuming idealised measurements. Our approach contrasts previous ones where only the dimension of the trusted party is known \cite{Moroder2016}. Notably, based only on a dimension assumption, there exists scenarios without any perfect devices \cite{Tavakoli2021, Tavakoli2018} in which unsteerable entanglement can be detected with similar resources as  in steering experiments  \cite{Piveteau2022, piveteau2023weak}. Moreover, knowledge of the dimension alone typically corresponds to known degrees of freedom but no detailed control over them. In contrast, our approach has the perspective that the experimenter controls the trusted device, but not flawlessly.

We study steering experiments where the trusted party's measurements feature small imprecision, i.e.~they nearly, but not exactly, correspond to the targeted measurements. We formalise the notion of measurement imprecision operationally, so that it can be  directly measured in the lab, or alternatively estimated from simulation of the experiment.  By considering concrete examples of well-known high-dimensional steering inequalities, we show that such small deviations from perfect control can have significant impact on the possibility of false positives. This leads us to address the main problem: if we are given a standard steering inequality, how can we determine the revision of the bound when the trusted party has imprecise measurements? We develop a general method for computing such bounds. It applies to any given bipartite steering inequality, in any dimension, with any given magnitude of imprecision. By design, it can be used also beyond  the standard steering task, to address imprecise measurements in the detection of more general entanglement properties in the steering scenario, for instance the Schmidt number \cite{Designolle2021, Designolle2022, Gois2023}. The result is analytical and the bounds are easily computable. Importantly, when applied to families of steering inequalities that are  commonly used for high-dimensional systems, we find that it can even provide exact bounds.

\textit{Steering inequalities.---} Consider two parties, Alice and Bob, who share an unknown bipartite quantum state $\rho$. Alice and Bob individually select private inputs $x$ and $y$ from pre-determined alphabets and produce outputs $a,b\in\{0,\ldots,d-1\}$ respectively. In a steering scenario, Alice's measurements, $\{A_{a|x}\}$, are unknown whereas Bob's target measurements are fully specified by $\{B^\text{targ}_{b|y}\}$. The quantum correlations, given by $p(a,b|x,y)=\Tr(A_{a|x}\otimes B^\text{targ}_{b|y}\rho)$, manifest steering if they admit no local hidden state (LHS) model. An LHS model takes the form $p(a,b|x,y)=\sum_\lambda p(\lambda)p(a|x,\lambda)\Tr\left(B^\text{targ}_{b|y}\sigma_\lambda\right)$, for some quantum states $\sigma_\lambda$. The set of correlations with an LHS model is identical to the set of quantum correlations attainable from separable states. 

Generally, steering can be detected through the violation of a steering inequality,
\begin{equation}\label{steerineq}
\mathcal{B}_0=\sum_{a,b,x,y} c_{abxy}\Tr\left(A_{a|x}\otimes B^\text{targ}_{b|y}\rho\right)\leq \beta_0^,
\end{equation}
where $c_{abxy}\geq 0$ are some coefficients and $\beta_0$ is a bound valid for any possible measurements for Alice and for any separable  $\rho$. Constructing such  inequalities is well-understood \cite{Cavalcanti2017}. In general, steering inequalities can also feature local terms for Alice, but we omit these cases in our presentation because they can, with small modifications, be treated analogously to our below results.

\textit{Imprecise measurements.---} We consider that Bob's lab measurements, $\{B_{b|y}\}$, nearly but not exactly correspond to his target measurements $\{B^\text{targ}_{b|y}\}$. To formalise this idea, we consider standard basis projections as the target measurements, i.e.~$\{B^\text{targ}_{b|y}\}$ are rank-one and projective. The set of possible lab measurements is qualitatively and quantitatively restricted. 

Qualitatively, the lab measurements are $d$-outcome POVMs which admit a decomposition in terms of a  classical mixture of standard basis measurements (rank-one and projective). A POVM $\{E_k\}$ admits such a model if it can be decomposed as $E_k=\sum_{\nu} p(\nu) N_k^{(\nu)}$, for some distribution $p(\nu)$ and some collection of rank-one projective measurements $\{N_k^{(\nu)}\}$. In this model, the measurements may be not only misaligned but also noisy, but the extremal measurements correspond to standard basis projections. The motivation for employing this class of measurements is two-fold. Firstly, higher-rank projective measurements are commonly implemented by resolving rank-one projections individually followed by post-processing (see e.g.~\cite{Ahrens2012, Ambrosio2014, Martinez2018, guo2023experimental}). Secondly, there exists non-projective measurements that cannot be simulated projectively \cite{Ariano2005} but their faithful implementation  requires the introduction of an ancilla system and  the use of control gates (see e.g.~\cite{Tavakoli2020, Smania2020, Martínez2023, feng2023higherdimensional}). This does arguably not correspond to an imprecision, but to a different experiment. 

Quantitatively,  we estimate how accurately the lab measurements approximate the target measurements by limiting the fidelity of each measurement operator,
\begin{equation}\label{fidcon}
\Tr\left(B_{b|y}B^\text{targ}_{b|y}\right)\geq 1-\epsilon_{by}.
\end{equation}
Here, $\epsilon_{by}\geq 0$ is the imprecision parameter associated to the $b$'th measurement operator for input $y$. Thus, $\epsilon=0$ corresponds to the standard steering scenario, with exactly known measurements. Typically, we will be interested in values of $\{\epsilon_{by}\}$ that are small but non-zero. Quantifying imprecisions via the fidelity has the advantage that the parameters become operationally meaningful; $\epsilon_{by}$ can be estimated by probing the lab measurement device with preparations of the eigenstate of the target measurement operator. 

We remark that our model of imprecise measurements is different from that proposed in \cite{Morelli2022} in the context of entanglement witnesses. There, the lab POVMs are not qualitatively restricted and the quantitative restriction is defined as an average fidelity over $b$. Our more conservative model is less detrimental for the analysis of realistic experimental data and arguably better captures imperfect lab measurements, while requiring the same modest resource cost.

\textit{Impact of imprecise measurements on steering.---}  We now investigate the significance of small imprecisions on relevant steering inequalities. That is, how significantly does a (tight) LHS bound $\beta_0$ in \eqref{steerineq} change in the presence of small deviations from the target measurement? For simplicity, we here consider that all measurement operators are equally imprecise, i.e.~$\epsilon=\epsilon_{by}$. We focus on a well-known family of high-dimensional steering inequalities \cite{Marciniak2015, Skrzypczyk2015}.

Alice and Bob have two inputs each and $d$ outputs. The steering inequality corresponds to the sum-total probability of correlated outcomes when the inputs are identical, i.e.~$c_{abxy}=\delta_{ab}\delta_{xy}$ in \eqref{steerineq}. The inequality reads $\mathcal{B}_0^{\text{mub}}= \sum_{a,x}\Tr\left(A_{a|x}\otimes B^\text{targ}_{a|x}\rho\right)\leq 1+\frac{1}{\sqrt{d}}=\beta_0^{\text{mub}}$. The bound is tight for all $d$. Here, the target measurements are mutually unbiased bases (MUBs), i.e.~they are rank-one, projective and satisfy $\forall b,b':\hspace{1mm}\Tr\left(B^\text{targ}_{b|1}B^\text{targ}_{b'|2}\right)=\frac{1}{d}$. Using a maximally entangled $d$-dimensional state and two MUBs for Alice, one obtains the largest quantum violation  $\mathcal{B}_0^{\text{mub}}=2$.

We now construct an explicit quantum strategy using a product state and imprecise measurements of the computational and Fourier bases. Choose $\rho=\ketbra{\psi_A}{\psi_A}\otimes \ketbra{\psi_B}{\psi_B}$ where $\ket{\psi_A}=\ket{0}$ and $\ket{\psi_B}=\nu\ket{0}+\sqrt{\frac{1-\nu^2}{d-1}}\left(\ket{1}+\ldots+\ket{d-1}\right)$, where
$\nu=\sqrt{(1+1/\sqrt{d})/2}$. Further, choose all Alice's measurements to be the computational basis $A_{a|x}=\ketbra{a}{a}$. 
To construct the imprecise measurements of Bob, define the ray $\ket{\phi}=\sqrt{1-\epsilon}\ket{0}+\sqrt{\frac{\epsilon}{d-1}}\left(\ket{1}+\ldots+\ket{d-1}\right)$ and choose the first outcome in the two bases as $B_{0|1}=\ketbra{\phi}{\phi}$ and $B_{0|2}=F\ketbra{\phi}{\phi}F^\dagger$ respectively, where $F=\frac{1}{\sqrt{d}}\sum_{j,k=0}^{d-1}e^{\frac{2\pi i}{d}jk}\ketbra{j}{k}$ is the Fourier transform. Clearly, the relations \eqref{fidcon} are saturated. The remaining measurement operators are irrelevant (since $p(a|x)=\delta_{a,0}$) and can be chosen in any compatible way. Writing $\mathcal{B}_\epsilon$ for steering functionals under imprecise measurements, we have  
\begin{align}\label{strategy}
\mathcal{B}_\epsilon^{\text{mub}}=\beta_0^{\text{mub}}+\frac{2}{\sqrt{d}}\left(\sqrt{\epsilon(1-\epsilon)(d-1)}-\epsilon\right), 
\end{align}
where the second term is a correction to the original LHS bound.  This model is valid for $\epsilon\leq \frac{1}{2}\left(1-\frac{1}{\sqrt{d}}\right)$, at the demarcation of which one finds the maximal quantum value $2$.

To  first order, the model scales as  $\mathcal{B}_\epsilon^{\text{mub}} \approx \beta_0^{\text{mub}} + 2\sqrt{\epsilon}\sqrt{\frac{d-1}{d}}$ which features a significant correction term which is relevant also for small $\epsilon$. To see this, we compute the share of the quantum violation gap that is erased due to the imprecise measurement, namely $\Delta= \frac{\mathcal{B}_\epsilon-\beta_0}{2-\beta_0}$. For instance, if the imprecision is $\epsilon=0.5\%$ in the qubit case, then we have $\mathcal{B}_\epsilon^{\text{mub}}\approx 1.80$, which corresponds to the sizable value $\Delta\approx 31\%$. Furthermore, for small $\epsilon$, we have  $\Delta\approx \frac{2\sqrt{\epsilon(d-1)}}{\sqrt{d}-1}$ which decreases monotically with $d$. In the limit $d\rightarrow \infty $ it becomes   $\Delta=2\sqrt{\epsilon(1-\epsilon)}$, which for $\epsilon=0.5\%$ is nevertheless still significant at around $14\%$. Importantly though,  one would realistically expect  $\epsilon$ to grow with the dimension. For example, for $d=100$, an imprecision of $\epsilon=2\%$ leads to a sizable $\Delta\approx 30\%$. Corrections on (at least) such magnitudes motivates the need to revise the LHS bound to explicitly take imprecisions into account.

\textit{Main result.---} We show how to systematically compute corrections to steering inequalities. Specifically, consider that we have a steering-type inequality \eqref{steerineq} and a set of imprecision parameters $\vec{\epsilon}=\{\epsilon_{by}\}$. We now expect that when optimised over all separable states, all measurements for Alice and all measurements of Bob with imprecision limited by $\vec{\epsilon}$, we will obtain a new  steering inequality, $\mathcal{B}_{\vec{\epsilon}}\leq \beta_{\vec{\epsilon}}$. However, computing the optimal value of $\beta_{\vec{\epsilon}}$ is a difficult optimisation problem. Therefore, we introduce a method to efficiently bound it from above.  To achieve this, we develop the following lemma, which may also be of independent interest.
\begin{lemma}\label{Lemma}
Let $\mathcal{S}_\epsilon(\ket{\psi})$ be the set of all states, in any dimension $D$, which are at least $\epsilon$-close in fidelity to $\ket{\psi}$. For every $\sigma\in \mathcal{S}_\epsilon(\ket{\psi})$ it holds that
\begin{equation}
\sigma \preceq \left(1+\mu\right)\ketbra{\psi}{\psi}+\frac{1}{2}\left(\sqrt{\mu^2+4\epsilon(1+\mu)}-\mu\right) \openone_D,
\end{equation}
for every choice of $\mu\geq -1$. 
\end{lemma}
The proof is given in Supplementary Material. Since our framework holds Bob's measurements as convex combinations of rank-one projections,  we can apply the lemma to every measurement operator $B_{b|y}$. Hence, for the evaluation of the $\vec{\epsilon}$-dependent LHS bound, we associate a parameter $\mu_{by}$ to each $B_{b|y}$, then use the lemma to  eliminate $\{B_{b|y}\}$ and then only deal with the known operators $\{B^{\text{targ}}_{b|y}\}$ and the identity. Let us define handy notations $z_{by}=\frac{1}{2}\left(\sqrt{\mu_{by}^2+4\epsilon_{by}(1+\mu_{by})}-\mu_{by}\right)$ and $\mu_\text{max}=\max_{b,y}\mu_{by}$. Then, we have
\begin{align}\nonumber 
	\mathcal{B}_{\vec{\epsilon}} &\leq \!\!\!\sum_{a,b,x,y}\!\!\! c_{abxy}\Tr\left(A_{a|x}\otimes \left((1+\mu_{by})\ketbra{\psi_{by}}{\psi_{by}}+z_{by}\openone\right)\rho\right)\\ \nonumber
	&\leq \left(1+\mu_\text{max}\right)\mathcal{B}_0 +\sum_{a,x} \Tr\left(A_{a|x}\rho_A\right)  \sum_{b,y} c_{abxy}z_{by}  \\\nonumber
	& \leq \left(1+\mu_\text{max}\right)\beta_0+\sum_{x} \max_{a} \left[\sum_{b,y} c_{abxy}z_{by} \right],
\end{align}
which holds for any choice of $\{\mu_{by}\}$. Furthermore, $z_{by}$ is non-negative and monotonically decreasing in $\mu_{by}$. Therefore the optimal choice of parameters, which minimises the right-hand-side, is to take all $\mu_{by}$ as large as possible, i.e.~$\mu_{by}=\mu_{\text{max}}$. We have arrived at our main result.
\begin{result}\label{result1}
	Consider any steering-type inequality \eqref{steerineq}, valid for some class of bipartite states $\rho$, with rank-one projective target measurements for Bob. If the lab measurements are associated with imprecision parameters $\vec{\epsilon}=\{\epsilon_{by}\}$, it holds that
	\begin{equation}\label{mainres}
		\mathcal{B}_{\vec{\epsilon}}\leq \min_{\mu\geq -1} \left(1+\mu\right)\beta_0+\frac{1}{2}\sum_{x} \max_{a} \Big(\sum_{b,y} c_{abxy}u_{by}\Big),
	\end{equation}
	where $u_{by}=\sqrt{\mu^2+4\epsilon_{by}(1+\mu)}-\mu$. 
\end{result}

The result has a number of noteworthy properties. Firstly, it is general. It applies to any linear steering inequality with arbitrary rank-one projective target measurements in any dimension. It can also take into account any set of imprecision parameters $\{\epsilon_{by}\}$.

Secondly, due to our separate use of Lemma~\ref{Lemma} for each measurement operator, the result does not require that $\{B_{b|y}\}$ satisfies $\sum_b B_{b|y}=\openone$.  This is a practically relevant situation, e.g.~when measuring high-dimensional optical systems. Many-outcome high-dimensional measurements are often implemented as a collection of $d$ distinct measurements, each projecting onto one specific basis element. Hence, the errors can occur separately, thus breaking the completeness of the intended measurement.

Thirdly, the bounds are obtained analytically, as every choice of the parameter $\mu\geq -1$ corresponds to a valid bound. Still, the best bound by this method, corresponding to the minimisation appearing in \eqref{mainres}, is favourably computed numerically. This is a simple numerical search since it only depends on a single real parameter. In the particularly appealing situation where all imprecision parameters are equal, $\epsilon=\epsilon_{by}$, the best bound admits a simple analytical expression,
\begin{equation}\label{ressimp}
	\mathcal{B}_{\vec{\epsilon}}\leq \beta_0 -\epsilon(2\beta_0-\chi)+2\sqrt{\epsilon(1-\epsilon)\beta_0(\chi-\beta_0)},
\end{equation}  
where  $\chi=\sum_{x} \max_{a} \left[\sum_{b,y} c_{abxy}\right]$. This is valid when $\epsilon\leq 1-\frac{\beta_0}{\chi}$. For small $\epsilon$, the first-order approximation is $\mathcal{B}_{\vec{\epsilon}} \lesssim\beta_0+2\sqrt{\beta_0(\chi-\beta_0)}\sqrt{\epsilon}$, scaling as the square-root of the imprecision.

Fourthly, Result~\ref{result1} makes no assumption on the bipartite state and therefore it applies to any inequality of the form \eqref{steerineq}. A natural example beyond standard steering is to limit the Schmidt number \cite{Terhal2000} of  $\rho$. The construction of such inequalities  is less straightforward but has  received recent  attention, again mainly using MUBs \cite{Designolle2021, Designolle2022, Gois2023}.

\textit{Case studies.---} The key question now is to evaluate the performance of Result~\ref{result1} in practice.  Interestingly, in spite of its generality, it can provide accurate and even exact bounds for well-known steering inequalities. We begin by showcasing this for the previously discussed high-dimensional tight steering inequality tailored for a pair of target MUBs,  $\mathcal{B}_0^{\text{mub}}\leq \beta_0^{\text{mub}}$. Such MUB-based criteria are particularly relevant, both because they are known to apply to arbitrary-dimensional systems and because of the several recent experiments using them, or variations of them, to detect steerability \cite{Zeng2018, Wang2018, Designolle2021, Srivastav2022, Qu2022, Qu22b}.

For the case of two MUBs in dimension $d$, let us first take all imprecision parameters to be equal. Then, the evaluation of Eq.~\eqref{ressimp} can be simplified to the expression in \eqref{strategy} previously obtained using an explicit quantum model with imprecise measurements. We thus conclude that for any $\epsilon$ and any dimension $d$, we obtain an exact bound on this steering inequality. Going further, consider the use of more than two MUBs; now  $c_{abxy}=\delta_{ab}\delta_{xy}$ in \eqref{steerineq} but with $n$ target MUBs ($x,y=1,\ldots,n$). An LHS bound is  $\frac{n}{d}(1+(d-1)/\sqrt{n})$ \cite{Skrzypczyk2015, Marciniak2015} but it is not exact in general.  The exact bound can, however, be computed for specific $(n,d)$. Therefore, to assess the accuracy of our result, we consider using three MUBs with $d=3,7,11$ and explicitly compute the exact standard steering bound. Then, we use this value of $\beta_0$ in \eqref{ressimp} to evaluate the $\epsilon$-dependent LHS bound. Comparing this result with those obtained from numerical search over quantum models with $n=3$ imprecise MUBs, we systematically match the analytical bound up to at least five decimal places. This suggests that Result~\ref{result1} can give exact results also  for several MUBs in high-dimensions.

\begin{figure}[t!]
	\centering
	\includegraphics[width=0.9\columnwidth]{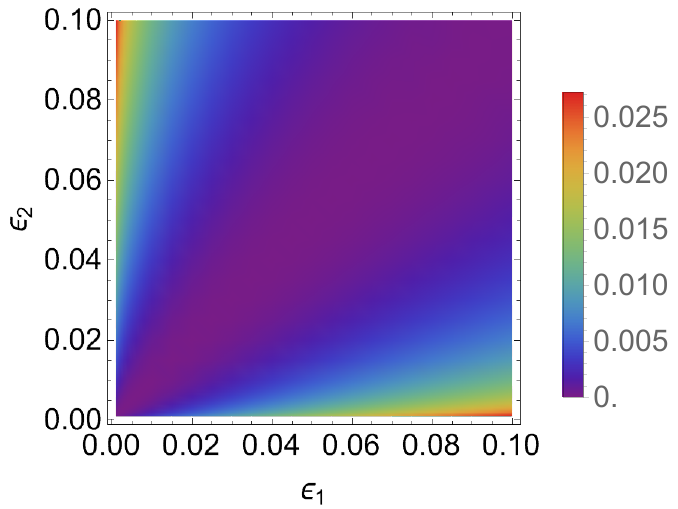}
	\caption{Accuracy of the steering bound under imprecise measurements for $\mathcal{B}_{\epsilon_1,\epsilon_2}^{\text{mub}}$ in dimension $7$. Each basis for the trusted party is attributed a distinct imprecision parameter. The plot shows the difference between the upper bound, obtained from \eqref{mainres}, and the best numerical lower bound for unsteerable states.}\label{Figd7}
\end{figure}

Next, let us focus on the more realistic situation in which different target measurements are attributed non-identical imprecision parameters. Indeed, more often than not, experimental control varies over complementary bases. For example, in Ref.~\cite{cao2023genuine} the average imprecision in the Pauli $Z$-basis observable is around $ 3\times 10^{-4}$, whereas in the $X$-basis it is twice as large and in the $Y$-basis it is nearly ten times as large. In view of this,  consider two MUBs and associate $\epsilon_1$ and $\epsilon_2$ to each of the bases respectively. To investigate the accuracy of the bound obtained from \eqref{mainres}, we have made a  $28\times 28$ non-uniform grid of the $(\epsilon_1,\epsilon_2)$-plane in the region $0\leq (\epsilon_1,\epsilon_2)\leq \frac{1}{10}$, and numerically optimised  $\mathcal{B}_{\epsilon_1,\epsilon_2}^{\text{mub}}$ over unsteerable strategies. Hence, for each point in the grid, we obtain a lower bound on the exact steering bound under imprecise measurements. Then, we have computed the upper bound on  $\mathcal{B}_{\epsilon_1,\epsilon_2}^{\text{mub}}$ from Eq.~\eqref{mainres} and taken the difference between the upper and lower bounds. This process was carried out separately for $d=3,7,11$. The case of $d=7$ is illustrated in Fig~\ref{Figd7}, with the other two cases behaving qualitatively similar. For all three choices of $d$, we observe that the bound is always close to exact, and that it tends to become more precise for comparable values of $(\epsilon_1,\epsilon_2)$. For instance, for $d=7$, in order for the upper bound to deviate by a magnitude of just $0.01$ from the lower bound when, say, $\epsilon_1=1\%$, we must choose the second imprecision parameter as large as $\epsilon\approx 9\%$.  Interestingly, as we increase $d$, we observe that the bound tends to become even more precise at a fixed value of $(\epsilon_1,\epsilon_2)$.

\textit{Application to experiments.---} In order to take imprecise measurements into account in a given test of steering, the experimenter needs to (i) estimate the imprecision parameters $\{\epsilon_{by}\}$, (ii) calculate the corrected steering bound via Result~\ref{result1}, and (iii) compare the corrected steering bound with the experimental estimate of the witness.  Step (i)  entails gathering a modest amount of additional data (unless $\{\epsilon_{by}\}$ is estimted via simulation), but some level of such  additional characterisation is necessary also in idealised steering experiments to evidence the assumption of exact knowledge of the measurement. In this way, the  procedure can be applied to strengthen many already performed quantum steering experiments. This is likely to be particularly relevant for experiments using high-dimensional systems, as this commonly comes with reduced precision on the quantum operations.

\textit{A witness robust to imprecisions.---} We conclude with an interesting observation: there exists nontrivial steering witnesses that, for sufficiently small but non-zero imprecision, require no correction to the steering bound.  To showcase this, we compare two different qubit steering inequalities.  On the one hand, we consider the steering inequality of Ref.~\cite{Saunders2010}, $\mathcal{B}_0^{\text{steer}}=\expect{A_1\otimes \sigma_X}+\expect{A_2\otimes \sigma_Z}+\expect{A_3\otimes\sigma_Y}\leq \sqrt{3}$. For Werner states, $\rho_v=v\ketbra{\psi^-}{\psi^-}+\frac{1-v}{4}\openone$, where $\ket{\psi^-}$ is the singlet, it detects steering when $v>\frac{1}{\sqrt{3}}$. On the other hand, we consider the so-called Elegant Bell inequality, where Alice optimally performs four measurements forming a tetrahedron on the Bloch sphere and Bob performs three measurements $B_1=\sigma_X$, $B_2=\sigma_Y$ and $B_3=\sigma_Z$ \cite{gisin2007bell, Tavakoli2020platonicsolids}. When assuming Bob's measurements, the resulting LHS-bound on the Bell quantity  becomes  $\mathcal{B}_0^{\text{bell}}=\sum_{x=1,2,3,4}\sum_{y=1,2,3} (-1)^{T_{xy}}\expect{A_x\otimes B_y}\leq 4$, where $T=\{[0,0,0],[0,1,1],[1,0,1],[1,1,0]\}$. As the best quantum strategy reaches $\mathcal{B}_0^{\text{bell}}=4\sqrt{3}$, this steering inequality also detects every Werner state with $v>\frac{1}{\sqrt{3}}$. 

We have numerically investigated how the initially equal detection ability of the two steering inequalities scales with $\epsilon$. We find $\mathcal{B}_\epsilon^{\text{steer}}=\sqrt{3}+2\sqrt{6}\sqrt{\epsilon(1-\epsilon)}-2\sqrt{3}\epsilon$, which is exact because it is reproduced by the upper bound obtained from \eqref{ressimp}. In sharp contrast, for the Bell-based steering inequality, we are unable to find any increase in the LHS bound when $\epsilon\lesssim 0.00326$, i.e.~the imprecise measurements appear to have no influence at all. This was repeatedly found using several different numerical optimisation algorithms. Consequently, the most relevant detrimental influence, namely when $\epsilon$ is very small, is eliminated. Above this threshold, we have $\mathcal{B}_\epsilon^{\text{bell}}=2\sqrt{3}+4\sqrt{6}\sqrt{\epsilon(1-\epsilon)}-4\sqrt{3}\epsilon$. One finds that the initially identical visibility threshold for detecting the Werner state deteriorates much less rapidly with $\epsilon$ in the case of $\mathcal{B}_\epsilon^{\text{bell}}$. This points to the interesting problem of designing steering inequalities that display such a plateau for small imprecision in the measurements.

\textit{Outlook.---} We have introduced quantum steering scenarios where the trusted party's measurements are not flawlessly controlled. We found that small imprecision can have large impacts on steering tests and showed how this can be formalised and systematically corrected. With realistic estimates of the imprecision parameters, our method and case studies can be applied to remove the assumption of idealised devices from quantum steering experiments.

It is likely that our approach, based on Lemma~\ref{Lemma}, could be applicable also for  multipartite steering scenarios, especially when just one party is trusted. Similarly, the approach might be relevant for eliminating the assumption of perfect quantum inputs in measurement-device-independent scenarios \cite{Kocsis2015, Zhao2020}. Furthermore, the Lemma can also be used to address the problem of entanglement witnessing in the presence of imprecise measurements. This problem was put forward in \cite{Rosset2012} and further developed in \cite{Morelli2022}. However, at least a naive attempt seems to lead to bounds that are  far from being exact. Developing a useful method of similar generality for the entanglement witness scenario remains a central open problem.

\textit{Note added.--- See also the related work \cite{Sarkar}.}

\begin{acknowledgments}
The author thanks Simon Morelli, Pharnam Bakhshinezhad, Subhayan Sarkar, Hayata Yamasaki, Roope Uola and Gabriele Cobucci for inspiring and supportive discussions. This work is supported by the Wenner-Gren Foundation and  by the Knut and Alice Wallenberg Foundation through the Wallenberg Center for Quantum Technology (WACQT).
\end{acknowledgments}

\bibliography{references_analytical_corrections}

\begin{thebibliography}{45}%
\makeatletter
\providecommand \@ifxundefined [1]{%
 \@ifx{#1\undefined}
}%
\providecommand \@ifnum [1]{%
 \ifnum #1\expandafter \@firstoftwo
 \else \expandafter \@secondoftwo
 \fi
}%
\providecommand \@ifx [1]{%
 \ifx #1\expandafter \@firstoftwo
 \else \expandafter \@secondoftwo
 \fi
}%
\providecommand \natexlab [1]{#1}%
\providecommand \enquote  [1]{``#1''}%
\providecommand \bibnamefont  [1]{#1}%
\providecommand \bibfnamefont [1]{#1}%
\providecommand \citenamefont [1]{#1}%
\providecommand \href@noop [0]{\@secondoftwo}%
\providecommand \href [0]{\begingroup \@sanitize@url \@href}%
\providecommand \@href[1]{\@@startlink{#1}\@@href}%
\providecommand \@@href[1]{\endgroup#1\@@endlink}%
\providecommand \@sanitize@url [0]{\catcode `\\12\catcode `\$12\catcode
  `\&12\catcode `\#12\catcode `\^12\catcode `\_12\catcode `\%12\relax}%
\providecommand \@@startlink[1]{}%
\providecommand \@@endlink[0]{}%
\providecommand \url  [0]{\begingroup\@sanitize@url \@url }%
\providecommand \@url [1]{\endgroup\@href {#1}{\urlprefix }}%
\providecommand \urlprefix  [0]{URL }%
\providecommand \Eprint [0]{\href }%
\providecommand \doibase [0]{https://doi.org/}%
\providecommand \selectlanguage [0]{\@gobble}%
\providecommand \bibinfo  [0]{\@secondoftwo}%
\providecommand \bibfield  [0]{\@secondoftwo}%
\providecommand \translation [1]{[#1]}%
\providecommand \BibitemOpen [0]{}%
\providecommand \bibitemStop [0]{}%
\providecommand \bibitemNoStop [0]{.\EOS\space}%
\providecommand \EOS [0]{\spacefactor3000\relax}%
\providecommand \BibitemShut  [1]{\csname bibitem#1\endcsname}%
\let\auto@bib@innerbib\@empty
\bibitem [{\citenamefont {Gühne}\ and\ \citenamefont
  {Tóth}(2009)}]{Guhne2009}%
  \BibitemOpen
  \bibfield  {author} {\bibinfo {author} {\bibfnamefont {O.}~\bibnamefont
  {Gühne}}\ and\ \bibinfo {author} {\bibfnamefont {G.}~\bibnamefont {Tóth}},\
  }\bibfield  {title} {\bibinfo {title} {Entanglement detection},\ }\href
  {https://doi.org/https://doi.org/10.1016/j.physrep.2009.02.004} {\bibfield
  {journal} {\bibinfo  {journal} {Physics Reports}\ }\textbf {\bibinfo {volume}
  {474}},\ \bibinfo {pages} {1} (\bibinfo {year} {2009})}\BibitemShut {NoStop}%
\bibitem [{\citenamefont {Brunner}\ \emph {et~al.}(2014)\citenamefont
  {Brunner}, \citenamefont {Cavalcanti}, \citenamefont {Pironio}, \citenamefont
  {Scarani},\ and\ \citenamefont {Wehner}}]{Brunner2014}%
  \BibitemOpen
  \bibfield  {author} {\bibinfo {author} {\bibfnamefont {N.}~\bibnamefont
  {Brunner}}, \bibinfo {author} {\bibfnamefont {D.}~\bibnamefont {Cavalcanti}},
  \bibinfo {author} {\bibfnamefont {S.}~\bibnamefont {Pironio}}, \bibinfo
  {author} {\bibfnamefont {V.}~\bibnamefont {Scarani}},\ and\ \bibinfo {author}
  {\bibfnamefont {S.}~\bibnamefont {Wehner}},\ }\bibfield  {title} {\bibinfo
  {title} {Bell nonlocality},\ }\href
  {https://doi.org/10.1103/RevModPhys.86.419} {\bibfield  {journal} {\bibinfo
  {journal} {Rev. Mod. Phys.}\ }\textbf {\bibinfo {volume} {86}},\ \bibinfo
  {pages} {419} (\bibinfo {year} {2014})}\BibitemShut {NoStop}%
\bibitem [{\citenamefont {Tavakoli}\ \emph {et~al.}(2022)\citenamefont
  {Tavakoli}, \citenamefont {Pozas-Kerstjens}, \citenamefont {Luo},\ and\
  \citenamefont {Renou}}]{Networks}%
  \BibitemOpen
  \bibfield  {author} {\bibinfo {author} {\bibfnamefont {A.}~\bibnamefont
  {Tavakoli}}, \bibinfo {author} {\bibfnamefont {A.}~\bibnamefont
  {Pozas-Kerstjens}}, \bibinfo {author} {\bibfnamefont {M.-X.}\ \bibnamefont
  {Luo}},\ and\ \bibinfo {author} {\bibfnamefont {M.-O.}\ \bibnamefont
  {Renou}},\ }\bibfield  {title} {\bibinfo {title} {Bell nonlocality in
  networks},\ }\href {https://doi.org/10.1088/1361-6633/ac41bb} {\bibfield
  {journal} {\bibinfo  {journal} {Reports on Progress in Physics}\ }\textbf
  {\bibinfo {volume} {85}},\ \bibinfo {pages} {056001} (\bibinfo {year}
  {2022})}\BibitemShut {NoStop}%
\bibitem [{\citenamefont {Augusiak}\ \emph {et~al.}(2014)\citenamefont
  {Augusiak}, \citenamefont {Demianowicz},\ and\ \citenamefont
  {Acín}}]{Augusiak2014}%
  \BibitemOpen
  \bibfield  {author} {\bibinfo {author} {\bibfnamefont {R.}~\bibnamefont
  {Augusiak}}, \bibinfo {author} {\bibfnamefont {M.}~\bibnamefont
  {Demianowicz}},\ and\ \bibinfo {author} {\bibfnamefont {A.}~\bibnamefont
  {Acín}},\ }\bibfield  {title} {\bibinfo {title} {Local hidden–variable
  models for entangled quantum states},\ }\href
  {https://doi.org/10.1088/1751-8113/47/42/424002} {\bibfield  {journal}
  {\bibinfo  {journal} {Journal of Physics A: Mathematical and Theoretical}\
  }\textbf {\bibinfo {volume} {47}},\ \bibinfo {pages} {424002} (\bibinfo
  {year} {2014})}\BibitemShut {NoStop}%
\bibitem [{\citenamefont {Uola}\ \emph {et~al.}(2020)\citenamefont {Uola},
  \citenamefont {Costa}, \citenamefont {Nguyen},\ and\ \citenamefont
  {G\"uhne}}]{Uola2020}%
  \BibitemOpen
  \bibfield  {author} {\bibinfo {author} {\bibfnamefont {R.}~\bibnamefont
  {Uola}}, \bibinfo {author} {\bibfnamefont {A.~C.~S.}\ \bibnamefont {Costa}},
  \bibinfo {author} {\bibfnamefont {H.~C.}\ \bibnamefont {Nguyen}},\ and\
  \bibinfo {author} {\bibfnamefont {O.}~\bibnamefont {G\"uhne}},\ }\bibfield
  {title} {\bibinfo {title} {Quantum steering},\ }\href
  {https://doi.org/10.1103/RevModPhys.92.015001} {\bibfield  {journal}
  {\bibinfo  {journal} {Rev. Mod. Phys.}\ }\textbf {\bibinfo {volume} {92}},\
  \bibinfo {pages} {015001} (\bibinfo {year} {2020})}\BibitemShut {NoStop}%
\bibitem [{\citenamefont {Saunders}\ \emph {et~al.}(2010)\citenamefont
  {Saunders}, \citenamefont {Jones}, \citenamefont {Wiseman},\ and\
  \citenamefont {Pryde}}]{Saunders2010}%
  \BibitemOpen
  \bibfield  {author} {\bibinfo {author} {\bibfnamefont {D.~J.}\ \bibnamefont
  {Saunders}}, \bibinfo {author} {\bibfnamefont {S.~J.}\ \bibnamefont {Jones}},
  \bibinfo {author} {\bibfnamefont {H.~M.}\ \bibnamefont {Wiseman}},\ and\
  \bibinfo {author} {\bibfnamefont {G.~J.}\ \bibnamefont {Pryde}},\ }\bibfield
  {title} {\bibinfo {title} {Experimental epr-steering using bell-local
  states},\ }\href {https://doi.org/10.1038/nphys1766} {\bibfield  {journal}
  {\bibinfo  {journal} {Nature Physics}\ }\textbf {\bibinfo {volume} {6}},\
  \bibinfo {pages} {845} (\bibinfo {year} {2010})}\BibitemShut {NoStop}%
\bibitem [{\citenamefont {Wittmann}\ \emph {et~al.}(2012)\citenamefont
  {Wittmann}, \citenamefont {Ramelow}, \citenamefont {Steinlechner},
  \citenamefont {Langford}, \citenamefont {Brunner}, \citenamefont {Wiseman},
  \citenamefont {Ursin},\ and\ \citenamefont {Zeilinger}}]{Wittmann2012}%
  \BibitemOpen
  \bibfield  {author} {\bibinfo {author} {\bibfnamefont {B.}~\bibnamefont
  {Wittmann}}, \bibinfo {author} {\bibfnamefont {S.}~\bibnamefont {Ramelow}},
  \bibinfo {author} {\bibfnamefont {F.}~\bibnamefont {Steinlechner}}, \bibinfo
  {author} {\bibfnamefont {N.~K.}\ \bibnamefont {Langford}}, \bibinfo {author}
  {\bibfnamefont {N.}~\bibnamefont {Brunner}}, \bibinfo {author} {\bibfnamefont
  {H.~M.}\ \bibnamefont {Wiseman}}, \bibinfo {author} {\bibfnamefont
  {R.}~\bibnamefont {Ursin}},\ and\ \bibinfo {author} {\bibfnamefont
  {A.}~\bibnamefont {Zeilinger}},\ }\bibfield  {title} {\bibinfo {title}
  {Loophole-free einstein–podolsky–rosen experiment via quantum steering},\
  }\href {https://doi.org/10.1088/1367-2630/14/5/053030} {\bibfield  {journal}
  {\bibinfo  {journal} {New Journal of Physics}\ }\textbf {\bibinfo {volume}
  {14}},\ \bibinfo {pages} {053030} (\bibinfo {year} {2012})}\BibitemShut
  {NoStop}%
\bibitem [{\citenamefont {H{\"a}ndchen}\ \emph {et~al.}(2012)\citenamefont
  {H{\"a}ndchen}, \citenamefont {Eberle}, \citenamefont {Steinlechner},
  \citenamefont {Samblowski}, \citenamefont {Franz}, \citenamefont {Werner},\
  and\ \citenamefont {Schnabel}}]{Händchen2012}%
  \BibitemOpen
  \bibfield  {author} {\bibinfo {author} {\bibfnamefont {V.}~\bibnamefont
  {H{\"a}ndchen}}, \bibinfo {author} {\bibfnamefont {T.}~\bibnamefont
  {Eberle}}, \bibinfo {author} {\bibfnamefont {S.}~\bibnamefont
  {Steinlechner}}, \bibinfo {author} {\bibfnamefont {A.}~\bibnamefont
  {Samblowski}}, \bibinfo {author} {\bibfnamefont {T.}~\bibnamefont {Franz}},
  \bibinfo {author} {\bibfnamefont {R.~F.}\ \bibnamefont {Werner}},\ and\
  \bibinfo {author} {\bibfnamefont {R.}~\bibnamefont {Schnabel}},\ }\bibfield
  {title} {\bibinfo {title} {Observation of one-way einstein--podolsky--rosen
  steering},\ }\href {https://doi.org/10.1038/nphoton.2012.202} {\bibfield
  {journal} {\bibinfo  {journal} {Nature Photonics}\ }\textbf {\bibinfo
  {volume} {6}},\ \bibinfo {pages} {596} (\bibinfo {year} {2012})}\BibitemShut
  {NoStop}%
\bibitem [{\citenamefont {Armstrong}\ \emph {et~al.}(2015)\citenamefont
  {Armstrong}, \citenamefont {Wang}, \citenamefont {Teh}, \citenamefont {Gong},
  \citenamefont {He}, \citenamefont {Janousek}, \citenamefont {Bachor},
  \citenamefont {Reid},\ and\ \citenamefont {Lam}}]{Armstrong2015}%
  \BibitemOpen
  \bibfield  {author} {\bibinfo {author} {\bibfnamefont {S.}~\bibnamefont
  {Armstrong}}, \bibinfo {author} {\bibfnamefont {M.}~\bibnamefont {Wang}},
  \bibinfo {author} {\bibfnamefont {R.~Y.}\ \bibnamefont {Teh}}, \bibinfo
  {author} {\bibfnamefont {Q.}~\bibnamefont {Gong}}, \bibinfo {author}
  {\bibfnamefont {Q.}~\bibnamefont {He}}, \bibinfo {author} {\bibfnamefont
  {J.}~\bibnamefont {Janousek}}, \bibinfo {author} {\bibfnamefont {H.-A.}\
  \bibnamefont {Bachor}}, \bibinfo {author} {\bibfnamefont {M.~D.}\
  \bibnamefont {Reid}},\ and\ \bibinfo {author} {\bibfnamefont {P.~K.}\
  \bibnamefont {Lam}},\ }\bibfield  {title} {\bibinfo {title} {Multipartite
  einstein--podolsky--rosen steering and genuine tripartite entanglement with
  optical networks},\ }\href {https://doi.org/10.1038/nphys3202} {\bibfield
  {journal} {\bibinfo  {journal} {Nature Physics}\ }\textbf {\bibinfo {volume}
  {11}},\ \bibinfo {pages} {167} (\bibinfo {year} {2015})}\BibitemShut
  {NoStop}%
\bibitem [{\citenamefont {Sun}\ \emph {et~al.}(2016)\citenamefont {Sun},
  \citenamefont {Ye}, \citenamefont {Xu}, \citenamefont {Xu}, \citenamefont
  {Tang}, \citenamefont {Wu}, \citenamefont {Chen}, \citenamefont {Li},\ and\
  \citenamefont {Guo}}]{Sun2016}%
  \BibitemOpen
  \bibfield  {author} {\bibinfo {author} {\bibfnamefont {K.}~\bibnamefont
  {Sun}}, \bibinfo {author} {\bibfnamefont {X.-J.}\ \bibnamefont {Ye}},
  \bibinfo {author} {\bibfnamefont {J.-S.}\ \bibnamefont {Xu}}, \bibinfo
  {author} {\bibfnamefont {X.-Y.}\ \bibnamefont {Xu}}, \bibinfo {author}
  {\bibfnamefont {J.-S.}\ \bibnamefont {Tang}}, \bibinfo {author}
  {\bibfnamefont {Y.-C.}\ \bibnamefont {Wu}}, \bibinfo {author} {\bibfnamefont
  {J.-L.}\ \bibnamefont {Chen}}, \bibinfo {author} {\bibfnamefont {C.-F.}\
  \bibnamefont {Li}},\ and\ \bibinfo {author} {\bibfnamefont {G.-C.}\
  \bibnamefont {Guo}},\ }\bibfield  {title} {\bibinfo {title} {Experimental
  quantification of asymmetric einstein-podolsky-rosen steering},\ }\href
  {https://doi.org/10.1103/PhysRevLett.116.160404} {\bibfield  {journal}
  {\bibinfo  {journal} {Phys. Rev. Lett.}\ }\textbf {\bibinfo {volume} {116}},\
  \bibinfo {pages} {160404} (\bibinfo {year} {2016})}\BibitemShut {NoStop}%
\bibitem [{\citenamefont {Zeng}\ \emph {et~al.}(2018)\citenamefont {Zeng},
  \citenamefont {Wang}, \citenamefont {Li},\ and\ \citenamefont
  {Zhang}}]{Zeng2018}%
  \BibitemOpen
  \bibfield  {author} {\bibinfo {author} {\bibfnamefont {Q.}~\bibnamefont
  {Zeng}}, \bibinfo {author} {\bibfnamefont {B.}~\bibnamefont {Wang}}, \bibinfo
  {author} {\bibfnamefont {P.}~\bibnamefont {Li}},\ and\ \bibinfo {author}
  {\bibfnamefont {X.}~\bibnamefont {Zhang}},\ }\bibfield  {title} {\bibinfo
  {title} {Experimental high-dimensional einstein-podolsky-rosen steering},\
  }\href {https://doi.org/10.1103/PhysRevLett.120.030401} {\bibfield  {journal}
  {\bibinfo  {journal} {Phys. Rev. Lett.}\ }\textbf {\bibinfo {volume} {120}},\
  \bibinfo {pages} {030401} (\bibinfo {year} {2018})}\BibitemShut {NoStop}%
\bibitem [{\citenamefont {Wang}\ \emph {et~al.}(2018)\citenamefont {Wang},
  \citenamefont {Paesani}, \citenamefont {Ding}, \citenamefont {Santagati},
  \citenamefont {Skrzypczyk}, \citenamefont {Salavrakos}, \citenamefont {Tura},
  \citenamefont {Augusiak}, \citenamefont {Mančinska}, \citenamefont {Bacco},
  \citenamefont {Bonneau}, \citenamefont {Silverstone}, \citenamefont {Gong},
  \citenamefont {Acín}, \citenamefont {Rottwitt}, \citenamefont {Oxenløwe},
  \citenamefont {O’Brien}, \citenamefont {Laing},\ and\ \citenamefont
  {Thompson}}]{Wang2018}%
  \BibitemOpen
  \bibfield  {author} {\bibinfo {author} {\bibfnamefont {J.}~\bibnamefont
  {Wang}}, \bibinfo {author} {\bibfnamefont {S.}~\bibnamefont {Paesani}},
  \bibinfo {author} {\bibfnamefont {Y.}~\bibnamefont {Ding}}, \bibinfo {author}
  {\bibfnamefont {R.}~\bibnamefont {Santagati}}, \bibinfo {author}
  {\bibfnamefont {P.}~\bibnamefont {Skrzypczyk}}, \bibinfo {author}
  {\bibfnamefont {A.}~\bibnamefont {Salavrakos}}, \bibinfo {author}
  {\bibfnamefont {J.}~\bibnamefont {Tura}}, \bibinfo {author} {\bibfnamefont
  {R.}~\bibnamefont {Augusiak}}, \bibinfo {author} {\bibfnamefont
  {L.}~\bibnamefont {Mančinska}}, \bibinfo {author} {\bibfnamefont
  {D.}~\bibnamefont {Bacco}}, \bibinfo {author} {\bibfnamefont
  {D.}~\bibnamefont {Bonneau}}, \bibinfo {author} {\bibfnamefont {J.~W.}\
  \bibnamefont {Silverstone}}, \bibinfo {author} {\bibfnamefont
  {Q.}~\bibnamefont {Gong}}, \bibinfo {author} {\bibfnamefont {A.}~\bibnamefont
  {Acín}}, \bibinfo {author} {\bibfnamefont {K.}~\bibnamefont {Rottwitt}},
  \bibinfo {author} {\bibfnamefont {L.~K.}\ \bibnamefont {Oxenløwe}}, \bibinfo
  {author} {\bibfnamefont {J.~L.}\ \bibnamefont {O’Brien}}, \bibinfo {author}
  {\bibfnamefont {A.}~\bibnamefont {Laing}},\ and\ \bibinfo {author}
  {\bibfnamefont {M.~G.}\ \bibnamefont {Thompson}},\ }\bibfield  {title}
  {\bibinfo {title} {Multidimensional quantum entanglement with large-scale
  integrated optics},\ }\href {https://doi.org/10.1126/science.aar7053}
  {\bibfield  {journal} {\bibinfo  {journal} {Science}\ }\textbf {\bibinfo
  {volume} {360}},\ \bibinfo {pages} {285} (\bibinfo {year} {2018})},\ \Eprint
  {https://arxiv.org/abs/https://www.science.org/doi/pdf/10.1126/science.aar7053}
  {https://www.science.org/doi/pdf/10.1126/science.aar7053} \BibitemShut
  {NoStop}%
\bibitem [{\citenamefont {Designolle}\ \emph {et~al.}(2021)\citenamefont
  {Designolle}, \citenamefont {Srivastav}, \citenamefont {Uola}, \citenamefont
  {Valencia}, \citenamefont {McCutcheon}, \citenamefont {Malik},\ and\
  \citenamefont {Brunner}}]{Designolle2021}%
  \BibitemOpen
  \bibfield  {author} {\bibinfo {author} {\bibfnamefont {S.}~\bibnamefont
  {Designolle}}, \bibinfo {author} {\bibfnamefont {V.}~\bibnamefont
  {Srivastav}}, \bibinfo {author} {\bibfnamefont {R.}~\bibnamefont {Uola}},
  \bibinfo {author} {\bibfnamefont {N.~H.}\ \bibnamefont {Valencia}}, \bibinfo
  {author} {\bibfnamefont {W.}~\bibnamefont {McCutcheon}}, \bibinfo {author}
  {\bibfnamefont {M.}~\bibnamefont {Malik}},\ and\ \bibinfo {author}
  {\bibfnamefont {N.}~\bibnamefont {Brunner}},\ }\bibfield  {title} {\bibinfo
  {title} {Genuine high-dimensional quantum steering},\ }\href
  {https://doi.org/10.1103/PhysRevLett.126.200404} {\bibfield  {journal}
  {\bibinfo  {journal} {Phys. Rev. Lett.}\ }\textbf {\bibinfo {volume} {126}},\
  \bibinfo {pages} {200404} (\bibinfo {year} {2021})}\BibitemShut {NoStop}%
\bibitem [{\citenamefont {Srivastav}\ \emph {et~al.}(2022)\citenamefont
  {Srivastav}, \citenamefont {Valencia}, \citenamefont {McCutcheon},
  \citenamefont {Leedumrongwatthanakun}, \citenamefont {Designolle},
  \citenamefont {Uola}, \citenamefont {Brunner},\ and\ \citenamefont
  {Malik}}]{Srivastav2022}%
  \BibitemOpen
  \bibfield  {author} {\bibinfo {author} {\bibfnamefont {V.}~\bibnamefont
  {Srivastav}}, \bibinfo {author} {\bibfnamefont {N.~H.}\ \bibnamefont
  {Valencia}}, \bibinfo {author} {\bibfnamefont {W.}~\bibnamefont
  {McCutcheon}}, \bibinfo {author} {\bibfnamefont {S.}~\bibnamefont
  {Leedumrongwatthanakun}}, \bibinfo {author} {\bibfnamefont {S.}~\bibnamefont
  {Designolle}}, \bibinfo {author} {\bibfnamefont {R.}~\bibnamefont {Uola}},
  \bibinfo {author} {\bibfnamefont {N.}~\bibnamefont {Brunner}},\ and\ \bibinfo
  {author} {\bibfnamefont {M.}~\bibnamefont {Malik}},\ }\bibfield  {title}
  {\bibinfo {title} {Quick quantum steering: Overcoming loss and noise with
  qudits},\ }\href {https://doi.org/10.1103/PhysRevX.12.041023} {\bibfield
  {journal} {\bibinfo  {journal} {Phys. Rev. X}\ }\textbf {\bibinfo {volume}
  {12}},\ \bibinfo {pages} {041023} (\bibinfo {year} {2022})}\BibitemShut
  {NoStop}%
\bibitem [{\citenamefont {Qu}\ \emph {et~al.}(2022{\natexlab{a}})\citenamefont
  {Qu}, \citenamefont {Wang}, \citenamefont {An}, \citenamefont {Wang},
  \citenamefont {Quan}, \citenamefont {Li}, \citenamefont {Gao}, \citenamefont
  {Li},\ and\ \citenamefont {Zhang}}]{Qu2022}%
  \BibitemOpen
  \bibfield  {author} {\bibinfo {author} {\bibfnamefont {R.}~\bibnamefont
  {Qu}}, \bibinfo {author} {\bibfnamefont {Y.}~\bibnamefont {Wang}}, \bibinfo
  {author} {\bibfnamefont {M.}~\bibnamefont {An}}, \bibinfo {author}
  {\bibfnamefont {F.}~\bibnamefont {Wang}}, \bibinfo {author} {\bibfnamefont
  {Q.}~\bibnamefont {Quan}}, \bibinfo {author} {\bibfnamefont {H.}~\bibnamefont
  {Li}}, \bibinfo {author} {\bibfnamefont {H.}~\bibnamefont {Gao}}, \bibinfo
  {author} {\bibfnamefont {F.}~\bibnamefont {Li}},\ and\ \bibinfo {author}
  {\bibfnamefont {P.}~\bibnamefont {Zhang}},\ }\bibfield  {title} {\bibinfo
  {title} {Retrieving high-dimensional quantum steering from a noisy
  environment with $n$ measurement settings},\ }\href
  {https://doi.org/10.1103/PhysRevLett.128.240402} {\bibfield  {journal}
  {\bibinfo  {journal} {Phys. Rev. Lett.}\ }\textbf {\bibinfo {volume} {128}},\
  \bibinfo {pages} {240402} (\bibinfo {year} {2022}{\natexlab{a}})}\BibitemShut
  {NoStop}%
\bibitem [{\citenamefont {Qu}\ \emph {et~al.}(2022{\natexlab{b}})\citenamefont
  {Qu}, \citenamefont {Wang}, \citenamefont {Zhang}, \citenamefont {Ru},
  \citenamefont {Wang}, \citenamefont {Gao}, \citenamefont {Li},\ and\
  \citenamefont {Zhang}}]{Qu22b}%
  \BibitemOpen
  \bibfield  {author} {\bibinfo {author} {\bibfnamefont {R.}~\bibnamefont
  {Qu}}, \bibinfo {author} {\bibfnamefont {Y.}~\bibnamefont {Wang}}, \bibinfo
  {author} {\bibfnamefont {X.}~\bibnamefont {Zhang}}, \bibinfo {author}
  {\bibfnamefont {S.}~\bibnamefont {Ru}}, \bibinfo {author} {\bibfnamefont
  {F.}~\bibnamefont {Wang}}, \bibinfo {author} {\bibfnamefont {H.}~\bibnamefont
  {Gao}}, \bibinfo {author} {\bibfnamefont {F.}~\bibnamefont {Li}},\ and\
  \bibinfo {author} {\bibfnamefont {P.}~\bibnamefont {Zhang}},\ }\bibfield
  {title} {\bibinfo {title} {Robust method for certifying genuine
  high-dimensional quantum steering with multimeasurement settings},\ }\href
  {https://doi.org/10.1364/OPTICA.454597} {\bibfield  {journal} {\bibinfo
  {journal} {Optica}\ }\textbf {\bibinfo {volume} {9}},\ \bibinfo {pages} {473}
  (\bibinfo {year} {2022}{\natexlab{b}})}\BibitemShut {NoStop}%
\bibitem [{\citenamefont {Huang}\ \emph {et~al.}(2021)\citenamefont {Huang},
  \citenamefont {Xiang}, \citenamefont {Guo}, \citenamefont {Wu}, \citenamefont
  {Liu}, \citenamefont {Li}, \citenamefont {Guo},\ and\ \citenamefont
  {Tavakoli}}]{Huang2021}%
  \BibitemOpen
  \bibfield  {author} {\bibinfo {author} {\bibfnamefont {C.-J.}\ \bibnamefont
  {Huang}}, \bibinfo {author} {\bibfnamefont {G.-Y.}\ \bibnamefont {Xiang}},
  \bibinfo {author} {\bibfnamefont {Y.}~\bibnamefont {Guo}}, \bibinfo {author}
  {\bibfnamefont {K.-D.}\ \bibnamefont {Wu}}, \bibinfo {author} {\bibfnamefont
  {B.-H.}\ \bibnamefont {Liu}}, \bibinfo {author} {\bibfnamefont {C.-F.}\
  \bibnamefont {Li}}, \bibinfo {author} {\bibfnamefont {G.-C.}\ \bibnamefont
  {Guo}},\ and\ \bibinfo {author} {\bibfnamefont {A.}~\bibnamefont
  {Tavakoli}},\ }\bibfield  {title} {\bibinfo {title} {Nonlocality, steering,
  and quantum state tomography in a single experiment},\ }\href
  {https://doi.org/10.1103/PhysRevLett.127.020401} {\bibfield  {journal}
  {\bibinfo  {journal} {Phys. Rev. Lett.}\ }\textbf {\bibinfo {volume} {127}},\
  \bibinfo {pages} {020401} (\bibinfo {year} {2021})}\BibitemShut {NoStop}%
\bibitem [{\citenamefont {Moroder}\ \emph {et~al.}(2016)\citenamefont
  {Moroder}, \citenamefont {Gittsovich}, \citenamefont {Huber}, \citenamefont
  {Uola},\ and\ \citenamefont {G\"uhne}}]{Moroder2016}%
  \BibitemOpen
  \bibfield  {author} {\bibinfo {author} {\bibfnamefont {T.}~\bibnamefont
  {Moroder}}, \bibinfo {author} {\bibfnamefont {O.}~\bibnamefont {Gittsovich}},
  \bibinfo {author} {\bibfnamefont {M.}~\bibnamefont {Huber}}, \bibinfo
  {author} {\bibfnamefont {R.}~\bibnamefont {Uola}},\ and\ \bibinfo {author}
  {\bibfnamefont {O.}~\bibnamefont {G\"uhne}},\ }\bibfield  {title} {\bibinfo
  {title} {Steering maps and their application to dimension-bounded steering},\
  }\href {https://doi.org/10.1103/PhysRevLett.116.090403} {\bibfield  {journal}
  {\bibinfo  {journal} {Phys. Rev. Lett.}\ }\textbf {\bibinfo {volume} {116}},\
  \bibinfo {pages} {090403} (\bibinfo {year} {2016})}\BibitemShut {NoStop}%
\bibitem [{\citenamefont {Tavakoli}\ \emph {et~al.}(2021)\citenamefont
  {Tavakoli}, \citenamefont {Pauwels}, \citenamefont {Woodhead},\ and\
  \citenamefont {Pironio}}]{Tavakoli2021}%
  \BibitemOpen
  \bibfield  {author} {\bibinfo {author} {\bibfnamefont {A.}~\bibnamefont
  {Tavakoli}}, \bibinfo {author} {\bibfnamefont {J.}~\bibnamefont {Pauwels}},
  \bibinfo {author} {\bibfnamefont {E.}~\bibnamefont {Woodhead}},\ and\
  \bibinfo {author} {\bibfnamefont {S.}~\bibnamefont {Pironio}},\ }\bibfield
  {title} {\bibinfo {title} {Correlations in entanglement-assisted
  prepare-and-measure scenarios},\ }\href
  {https://doi.org/10.1103/PRXQuantum.2.040357} {\bibfield  {journal} {\bibinfo
   {journal} {PRX Quantum}\ }\textbf {\bibinfo {volume} {2}},\ \bibinfo {pages}
  {040357} (\bibinfo {year} {2021})}\BibitemShut {NoStop}%
\bibitem [{\citenamefont {Tavakoli}\ \emph {et~al.}(2018)\citenamefont
  {Tavakoli}, \citenamefont {Abbott}, \citenamefont {Renou}, \citenamefont
  {Gisin},\ and\ \citenamefont {Brunner}}]{Tavakoli2018}%
  \BibitemOpen
  \bibfield  {author} {\bibinfo {author} {\bibfnamefont {A.}~\bibnamefont
  {Tavakoli}}, \bibinfo {author} {\bibfnamefont {A.~A.}\ \bibnamefont
  {Abbott}}, \bibinfo {author} {\bibfnamefont {M.-O.}\ \bibnamefont {Renou}},
  \bibinfo {author} {\bibfnamefont {N.}~\bibnamefont {Gisin}},\ and\ \bibinfo
  {author} {\bibfnamefont {N.}~\bibnamefont {Brunner}},\ }\bibfield  {title}
  {\bibinfo {title} {Semi-device-independent characterization of multipartite
  entanglement of states and measurements},\ }\href
  {https://doi.org/10.1103/PhysRevA.98.052333} {\bibfield  {journal} {\bibinfo
  {journal} {Phys. Rev. A}\ }\textbf {\bibinfo {volume} {98}},\ \bibinfo
  {pages} {052333} (\bibinfo {year} {2018})}\BibitemShut {NoStop}%
\bibitem [{\citenamefont {Piveteau}\ \emph {et~al.}(2022)\citenamefont
  {Piveteau}, \citenamefont {Pauwels}, \citenamefont {H{\aa}kansson},
  \citenamefont {Muhammad}, \citenamefont {Bourennane},\ and\ \citenamefont
  {Tavakoli}}]{Piveteau2022}%
  \BibitemOpen
  \bibfield  {author} {\bibinfo {author} {\bibfnamefont {A.}~\bibnamefont
  {Piveteau}}, \bibinfo {author} {\bibfnamefont {J.}~\bibnamefont {Pauwels}},
  \bibinfo {author} {\bibfnamefont {E.}~\bibnamefont {H{\aa}kansson}}, \bibinfo
  {author} {\bibfnamefont {S.}~\bibnamefont {Muhammad}}, \bibinfo {author}
  {\bibfnamefont {M.}~\bibnamefont {Bourennane}},\ and\ \bibinfo {author}
  {\bibfnamefont {A.}~\bibnamefont {Tavakoli}},\ }\bibfield  {title} {\bibinfo
  {title} {Entanglement-assisted quantum communication with simple
  measurements},\ }\href {https://doi.org/10.1038/s41467-022-33922-5}
  {\bibfield  {journal} {\bibinfo  {journal} {Nature Communications}\ }\textbf
  {\bibinfo {volume} {13}},\ \bibinfo {pages} {7878} (\bibinfo {year}
  {2022})}\BibitemShut {NoStop}%
\bibitem [{\citenamefont {Piveteau}\ \emph {et~al.}(2023)\citenamefont
  {Piveteau}, \citenamefont {Abbott}, \citenamefont {Muhammad}, \citenamefont
  {Bourennane},\ and\ \citenamefont {Tavakoli}}]{piveteau2023weak}%
  \BibitemOpen
  \bibfield  {author} {\bibinfo {author} {\bibfnamefont {A.}~\bibnamefont
  {Piveteau}}, \bibinfo {author} {\bibfnamefont {A.~A.}\ \bibnamefont
  {Abbott}}, \bibinfo {author} {\bibfnamefont {S.}~\bibnamefont {Muhammad}},
  \bibinfo {author} {\bibfnamefont {M.}~\bibnamefont {Bourennane}},\ and\
  \bibinfo {author} {\bibfnamefont {A.}~\bibnamefont {Tavakoli}},\ }\href@noop
  {} {\bibinfo {title} {Weak entanglement improves quantum communication using
  only passive linear optics}} (\bibinfo {year} {2023}),\ \Eprint
  {https://arxiv.org/abs/2303.07907} {arXiv:2303.07907 [quant-ph]} \BibitemShut
  {NoStop}%
\bibitem [{\citenamefont {Designolle}(2022)}]{Designolle2022}%
  \BibitemOpen
  \bibfield  {author} {\bibinfo {author} {\bibfnamefont {S.}~\bibnamefont
  {Designolle}},\ }\bibfield  {title} {\bibinfo {title} {Robust genuine
  high-dimensional steering with many measurements},\ }\href
  {https://doi.org/10.1103/PhysRevA.105.032430} {\bibfield  {journal} {\bibinfo
   {journal} {Phys. Rev. A}\ }\textbf {\bibinfo {volume} {105}},\ \bibinfo
  {pages} {032430} (\bibinfo {year} {2022})}\BibitemShut {NoStop}%
\bibitem [{\citenamefont {de~Gois}\ \emph {et~al.}(2023)\citenamefont
  {de~Gois}, \citenamefont {Pl\'avala}, \citenamefont {Schwonnek},\ and\
  \citenamefont {G\"uhne}}]{Gois2023}%
  \BibitemOpen
  \bibfield  {author} {\bibinfo {author} {\bibfnamefont {C.}~\bibnamefont
  {de~Gois}}, \bibinfo {author} {\bibfnamefont {M.}~\bibnamefont {Pl\'avala}},
  \bibinfo {author} {\bibfnamefont {R.}~\bibnamefont {Schwonnek}},\ and\
  \bibinfo {author} {\bibfnamefont {O.}~\bibnamefont {G\"uhne}},\ }\bibfield
  {title} {\bibinfo {title} {Complete hierarchy for high-dimensional steering
  certification},\ }\href {https://doi.org/10.1103/PhysRevLett.131.010201}
  {\bibfield  {journal} {\bibinfo  {journal} {Phys. Rev. Lett.}\ }\textbf
  {\bibinfo {volume} {131}},\ \bibinfo {pages} {010201} (\bibinfo {year}
  {2023})}\BibitemShut {NoStop}%
\bibitem [{\citenamefont {Cavalcanti}\ and\ \citenamefont
  {Skrzypczyk}(2016)}]{Cavalcanti2017}%
  \BibitemOpen
  \bibfield  {author} {\bibinfo {author} {\bibfnamefont {D.}~\bibnamefont
  {Cavalcanti}}\ and\ \bibinfo {author} {\bibfnamefont {P.}~\bibnamefont
  {Skrzypczyk}},\ }\bibfield  {title} {\bibinfo {title} {Quantum steering: a
  review with focus on semidefinite programming},\ }\href
  {https://doi.org/10.1088/1361-6633/80/2/024001} {\bibfield  {journal}
  {\bibinfo  {journal} {Reports on Progress in Physics}\ }\textbf {\bibinfo
  {volume} {80}},\ \bibinfo {pages} {024001} (\bibinfo {year}
  {2016})}\BibitemShut {NoStop}%
\bibitem [{\citenamefont {Ahrens}\ \emph {et~al.}(2012)\citenamefont {Ahrens},
  \citenamefont {Badziag}, \citenamefont {Cabello},\ and\ \citenamefont
  {Bourennane}}]{Ahrens2012}%
  \BibitemOpen
  \bibfield  {author} {\bibinfo {author} {\bibfnamefont {J.}~\bibnamefont
  {Ahrens}}, \bibinfo {author} {\bibfnamefont {P.}~\bibnamefont {Badziag}},
  \bibinfo {author} {\bibfnamefont {A.}~\bibnamefont {Cabello}},\ and\ \bibinfo
  {author} {\bibfnamefont {M.}~\bibnamefont {Bourennane}},\ }\bibfield  {title}
  {\bibinfo {title} {Experimental device-independent tests of classical and
  quantum dimensions},\ }\href {https://doi.org/10.1038/nphys2333} {\bibfield
  {journal} {\bibinfo  {journal} {Nature Physics}\ }\textbf {\bibinfo {volume}
  {8}},\ \bibinfo {pages} {592} (\bibinfo {year} {2012})}\BibitemShut {NoStop}%
\bibitem [{\citenamefont {D'Ambrosio}\ \emph {et~al.}(2014)\citenamefont
  {D'Ambrosio}, \citenamefont {Bisesto}, \citenamefont {Sciarrino},
  \citenamefont {Barra}, \citenamefont {Lima},\ and\ \citenamefont
  {Cabello}}]{Ambrosio2014}%
  \BibitemOpen
  \bibfield  {author} {\bibinfo {author} {\bibfnamefont {V.}~\bibnamefont
  {D'Ambrosio}}, \bibinfo {author} {\bibfnamefont {F.}~\bibnamefont {Bisesto}},
  \bibinfo {author} {\bibfnamefont {F.}~\bibnamefont {Sciarrino}}, \bibinfo
  {author} {\bibfnamefont {J.~F.}\ \bibnamefont {Barra}}, \bibinfo {author}
  {\bibfnamefont {G.}~\bibnamefont {Lima}},\ and\ \bibinfo {author}
  {\bibfnamefont {A.}~\bibnamefont {Cabello}},\ }\bibfield  {title} {\bibinfo
  {title} {Device-independent certification of high-dimensional quantum
  systems},\ }\href {https://doi.org/10.1103/PhysRevLett.112.140503} {\bibfield
   {journal} {\bibinfo  {journal} {Phys. Rev. Lett.}\ }\textbf {\bibinfo
  {volume} {112}},\ \bibinfo {pages} {140503} (\bibinfo {year}
  {2014})}\BibitemShut {NoStop}%
\bibitem [{\citenamefont {Mart\'{\i}nez}\ \emph {et~al.}(2018)\citenamefont
  {Mart\'{\i}nez}, \citenamefont {Tavakoli}, \citenamefont {Casanova},
  \citenamefont {Ca\~nas}, \citenamefont {Marques},\ and\ \citenamefont
  {Lima}}]{Martinez2018}%
  \BibitemOpen
  \bibfield  {author} {\bibinfo {author} {\bibfnamefont {D.}~\bibnamefont
  {Mart\'{\i}nez}}, \bibinfo {author} {\bibfnamefont {A.}~\bibnamefont
  {Tavakoli}}, \bibinfo {author} {\bibfnamefont {M.}~\bibnamefont {Casanova}},
  \bibinfo {author} {\bibfnamefont {G.}~\bibnamefont {Ca\~nas}}, \bibinfo
  {author} {\bibfnamefont {B.}~\bibnamefont {Marques}},\ and\ \bibinfo {author}
  {\bibfnamefont {G.}~\bibnamefont {Lima}},\ }\bibfield  {title} {\bibinfo
  {title} {High-dimensional quantum communication complexity beyond strategies
  based on bell's theorem},\ }\href
  {https://doi.org/10.1103/PhysRevLett.121.150504} {\bibfield  {journal}
  {\bibinfo  {journal} {Phys. Rev. Lett.}\ }\textbf {\bibinfo {volume} {121}},\
  \bibinfo {pages} {150504} (\bibinfo {year} {2018})}\BibitemShut {NoStop}%
\bibitem [{\citenamefont {Guo}\ \emph {et~al.}(2023)\citenamefont {Guo},
  \citenamefont {Tang}, \citenamefont {Pauwels}, \citenamefont {Cruzeiro},
  \citenamefont {Hu}, \citenamefont {Liu}, \citenamefont {Huang}, \citenamefont
  {Li}, \citenamefont {Guo},\ and\ \citenamefont
  {Tavakoli}}]{guo2023experimental}%
  \BibitemOpen
  \bibfield  {author} {\bibinfo {author} {\bibfnamefont {Y.}~\bibnamefont
  {Guo}}, \bibinfo {author} {\bibfnamefont {H.}~\bibnamefont {Tang}}, \bibinfo
  {author} {\bibfnamefont {J.}~\bibnamefont {Pauwels}}, \bibinfo {author}
  {\bibfnamefont {E.~Z.}\ \bibnamefont {Cruzeiro}}, \bibinfo {author}
  {\bibfnamefont {X.-M.}\ \bibnamefont {Hu}}, \bibinfo {author} {\bibfnamefont
  {B.-H.}\ \bibnamefont {Liu}}, \bibinfo {author} {\bibfnamefont {Y.-F.}\
  \bibnamefont {Huang}}, \bibinfo {author} {\bibfnamefont {C.-F.}\ \bibnamefont
  {Li}}, \bibinfo {author} {\bibfnamefont {G.-C.}\ \bibnamefont {Guo}},\ and\
  \bibinfo {author} {\bibfnamefont {A.}~\bibnamefont {Tavakoli}},\ }\href@noop
  {} {\bibinfo {title} {Experimental higher-dimensional entanglement advantage
  over qubit channel}} (\bibinfo {year} {2023}),\ \Eprint
  {https://arxiv.org/abs/2306.13495} {arXiv:2306.13495 [quant-ph]} \BibitemShut
  {NoStop}%
\bibitem [{\citenamefont {D'Ariano}\ \emph {et~al.}(2005)\citenamefont
  {D'Ariano}, \citenamefont {Presti},\ and\ \citenamefont
  {Perinotti}}]{Ariano2005}%
  \BibitemOpen
  \bibfield  {author} {\bibinfo {author} {\bibfnamefont {G.~M.}\ \bibnamefont
  {D'Ariano}}, \bibinfo {author} {\bibfnamefont {P.~L.}\ \bibnamefont
  {Presti}},\ and\ \bibinfo {author} {\bibfnamefont {P.}~\bibnamefont
  {Perinotti}},\ }\bibfield  {title} {\bibinfo {title} {Classical randomness in
  quantum measurements},\ }\href {https://doi.org/10.1088/0305-4470/38/26/010}
  {\bibfield  {journal} {\bibinfo  {journal} {Journal of Physics A:
  Mathematical and General}\ }\textbf {\bibinfo {volume} {38}},\ \bibinfo
  {pages} {5979} (\bibinfo {year} {2005})}\BibitemShut {NoStop}%
\bibitem [{\citenamefont {Tavakoli}\ \emph {et~al.}(2020)\citenamefont
  {Tavakoli}, \citenamefont {Smania}, \citenamefont {Vértesi}, \citenamefont
  {Brunner},\ and\ \citenamefont {Bourennane}}]{Tavakoli2020}%
  \BibitemOpen
  \bibfield  {author} {\bibinfo {author} {\bibfnamefont {A.}~\bibnamefont
  {Tavakoli}}, \bibinfo {author} {\bibfnamefont {M.}~\bibnamefont {Smania}},
  \bibinfo {author} {\bibfnamefont {T.}~\bibnamefont {Vértesi}}, \bibinfo
  {author} {\bibfnamefont {N.}~\bibnamefont {Brunner}},\ and\ \bibinfo {author}
  {\bibfnamefont {M.}~\bibnamefont {Bourennane}},\ }\bibfield  {title}
  {\bibinfo {title} {Self-testing nonprojective quantum measurements in
  prepare-and-measure experiments},\ }\href
  {https://doi.org/10.1126/sciadv.aaw6664} {\bibfield  {journal} {\bibinfo
  {journal} {Science Advances}\ }\textbf {\bibinfo {volume} {6}},\ \bibinfo
  {pages} {eaaw6664} (\bibinfo {year} {2020})},\ \Eprint
  {https://arxiv.org/abs/https://www.science.org/doi/pdf/10.1126/sciadv.aaw6664}
  {https://www.science.org/doi/pdf/10.1126/sciadv.aaw6664} \BibitemShut
  {NoStop}%
\bibitem [{\citenamefont {Smania}\ \emph {et~al.}(2020)\citenamefont {Smania},
  \citenamefont {Mironowicz}, \citenamefont {Nawareg}, \citenamefont
  {Paw{\l}owski}, \citenamefont {Cabello},\ and\ \citenamefont
  {Bourennane}}]{Smania2020}%
  \BibitemOpen
  \bibfield  {author} {\bibinfo {author} {\bibfnamefont {M.}~\bibnamefont
  {Smania}}, \bibinfo {author} {\bibfnamefont {P.}~\bibnamefont {Mironowicz}},
  \bibinfo {author} {\bibfnamefont {M.}~\bibnamefont {Nawareg}}, \bibinfo
  {author} {\bibfnamefont {M.}~\bibnamefont {Paw{\l}owski}}, \bibinfo {author}
  {\bibfnamefont {A.}~\bibnamefont {Cabello}},\ and\ \bibinfo {author}
  {\bibfnamefont {M.}~\bibnamefont {Bourennane}},\ }\bibfield  {title}
  {\bibinfo {title} {Experimental certification of an informationally complete
  quantum measurement in a device-independent protocol},\ }\href
  {https://doi.org/10.1364/OPTICA.377959} {\bibfield  {journal} {\bibinfo
  {journal} {Optica}\ }\textbf {\bibinfo {volume} {7}},\ \bibinfo {pages} {123}
  (\bibinfo {year} {2020})}\BibitemShut {NoStop}%
\bibitem [{\citenamefont {Mart{\'i}nez}\ \emph {et~al.}(2023)\citenamefont
  {Mart{\'i}nez}, \citenamefont {G{\'o}mez}, \citenamefont {Cari{\~{n}}e},
  \citenamefont {Pereira}, \citenamefont {Delgado}, \citenamefont {Walborn},
  \citenamefont {Tavakoli},\ and\ \citenamefont {Lima}}]{Martínez2023}%
  \BibitemOpen
  \bibfield  {author} {\bibinfo {author} {\bibfnamefont {D.}~\bibnamefont
  {Mart{\'i}nez}}, \bibinfo {author} {\bibfnamefont {E.~S.}\ \bibnamefont
  {G{\'o}mez}}, \bibinfo {author} {\bibfnamefont {J.}~\bibnamefont
  {Cari{\~{n}}e}}, \bibinfo {author} {\bibfnamefont {L.}~\bibnamefont
  {Pereira}}, \bibinfo {author} {\bibfnamefont {A.}~\bibnamefont {Delgado}},
  \bibinfo {author} {\bibfnamefont {S.~P.}\ \bibnamefont {Walborn}}, \bibinfo
  {author} {\bibfnamefont {A.}~\bibnamefont {Tavakoli}},\ and\ \bibinfo
  {author} {\bibfnamefont {G.}~\bibnamefont {Lima}},\ }\bibfield  {title}
  {\bibinfo {title} {Certification of a non-projective qudit measurement using
  multiport beamsplitters},\ }\href
  {https://doi.org/10.1038/s41567-022-01845-z} {\bibfield  {journal} {\bibinfo
  {journal} {Nature Physics}\ }\textbf {\bibinfo {volume} {19}},\ \bibinfo
  {pages} {190} (\bibinfo {year} {2023})}\BibitemShut {NoStop}%
\bibitem [{\citenamefont {Feng}\ \emph {et~al.}(2023)\citenamefont {Feng},
  \citenamefont {Hu}, \citenamefont {Zhang}, \citenamefont {Cheng},
  \citenamefont {Zhang}, \citenamefont {Guo}, \citenamefont {Ding},
  \citenamefont {Hou}, \citenamefont {Sun}, \citenamefont {Guo}, \citenamefont
  {Dai}, \citenamefont {Tavakoli}, \citenamefont {Ren},\ and\ \citenamefont
  {Liu}}]{feng2023higherdimensional}%
  \BibitemOpen
  \bibfield  {author} {\bibinfo {author} {\bibfnamefont {L.-T.}\ \bibnamefont
  {Feng}}, \bibinfo {author} {\bibfnamefont {X.-M.}\ \bibnamefont {Hu}},
  \bibinfo {author} {\bibfnamefont {M.}~\bibnamefont {Zhang}}, \bibinfo
  {author} {\bibfnamefont {Y.-J.}\ \bibnamefont {Cheng}}, \bibinfo {author}
  {\bibfnamefont {C.}~\bibnamefont {Zhang}}, \bibinfo {author} {\bibfnamefont
  {Y.}~\bibnamefont {Guo}}, \bibinfo {author} {\bibfnamefont {Y.-Y.}\
  \bibnamefont {Ding}}, \bibinfo {author} {\bibfnamefont {Z.}~\bibnamefont
  {Hou}}, \bibinfo {author} {\bibfnamefont {F.-W.}\ \bibnamefont {Sun}},
  \bibinfo {author} {\bibfnamefont {G.-C.}\ \bibnamefont {Guo}}, \bibinfo
  {author} {\bibfnamefont {D.-X.}\ \bibnamefont {Dai}}, \bibinfo {author}
  {\bibfnamefont {A.}~\bibnamefont {Tavakoli}}, \bibinfo {author}
  {\bibfnamefont {X.-F.}\ \bibnamefont {Ren}},\ and\ \bibinfo {author}
  {\bibfnamefont {B.-H.}\ \bibnamefont {Liu}},\ }\href@noop {} {\bibinfo
  {title} {Higher-dimensional symmetric informationally complete measurement
  via programmable photonic integrated optics}} (\bibinfo {year} {2023}),\
  \Eprint {https://arxiv.org/abs/2310.08838} {arXiv:2310.08838 [quant-ph]}
  \BibitemShut {NoStop}%
\bibitem [{\citenamefont {Morelli}\ \emph {et~al.}(2022)\citenamefont
  {Morelli}, \citenamefont {Yamasaki}, \citenamefont {Huber},\ and\
  \citenamefont {Tavakoli}}]{Morelli2022}%
  \BibitemOpen
  \bibfield  {author} {\bibinfo {author} {\bibfnamefont {S.}~\bibnamefont
  {Morelli}}, \bibinfo {author} {\bibfnamefont {H.}~\bibnamefont {Yamasaki}},
  \bibinfo {author} {\bibfnamefont {M.}~\bibnamefont {Huber}},\ and\ \bibinfo
  {author} {\bibfnamefont {A.}~\bibnamefont {Tavakoli}},\ }\bibfield  {title}
  {\bibinfo {title} {Entanglement detection with imprecise measurements},\
  }\href {https://doi.org/10.1103/PhysRevLett.128.250501} {\bibfield  {journal}
  {\bibinfo  {journal} {Phys. Rev. Lett.}\ }\textbf {\bibinfo {volume} {128}},\
  \bibinfo {pages} {250501} (\bibinfo {year} {2022})}\BibitemShut {NoStop}%
\bibitem [{\citenamefont {Marciniak}\ \emph {et~al.}(2015)\citenamefont
  {Marciniak}, \citenamefont {Rutkowski}, \citenamefont {Yin}, \citenamefont
  {Horodecki},\ and\ \citenamefont {Horodecki}}]{Marciniak2015}%
  \BibitemOpen
  \bibfield  {author} {\bibinfo {author} {\bibfnamefont {M.}~\bibnamefont
  {Marciniak}}, \bibinfo {author} {\bibfnamefont {A.}~\bibnamefont
  {Rutkowski}}, \bibinfo {author} {\bibfnamefont {Z.}~\bibnamefont {Yin}},
  \bibinfo {author} {\bibfnamefont {M.}~\bibnamefont {Horodecki}},\ and\
  \bibinfo {author} {\bibfnamefont {R.}~\bibnamefont {Horodecki}},\ }\bibfield
  {title} {\bibinfo {title} {Unbounded violation of quantum steering
  inequalities},\ }\href {https://doi.org/10.1103/PhysRevLett.115.170401}
  {\bibfield  {journal} {\bibinfo  {journal} {Phys. Rev. Lett.}\ }\textbf
  {\bibinfo {volume} {115}},\ \bibinfo {pages} {170401} (\bibinfo {year}
  {2015})}\BibitemShut {NoStop}%
\bibitem [{\citenamefont {Skrzypczyk}\ and\ \citenamefont
  {Cavalcanti}(2015)}]{Skrzypczyk2015}%
  \BibitemOpen
  \bibfield  {author} {\bibinfo {author} {\bibfnamefont {P.}~\bibnamefont
  {Skrzypczyk}}\ and\ \bibinfo {author} {\bibfnamefont {D.}~\bibnamefont
  {Cavalcanti}},\ }\bibfield  {title} {\bibinfo {title} {Loss-tolerant
  einstein-podolsky-rosen steering for arbitrary-dimensional states: Joint
  measurability and unbounded violations under losses},\ }\href
  {https://doi.org/10.1103/PhysRevA.92.022354} {\bibfield  {journal} {\bibinfo
  {journal} {Phys. Rev. A}\ }\textbf {\bibinfo {volume} {92}},\ \bibinfo
  {pages} {022354} (\bibinfo {year} {2015})}\BibitemShut {NoStop}%
\bibitem [{\citenamefont {Terhal}\ and\ \citenamefont
  {Horodecki}(2000)}]{Terhal2000}%
  \BibitemOpen
  \bibfield  {author} {\bibinfo {author} {\bibfnamefont {B.~M.}\ \bibnamefont
  {Terhal}}\ and\ \bibinfo {author} {\bibfnamefont {P.}~\bibnamefont
  {Horodecki}},\ }\bibfield  {title} {\bibinfo {title} {Schmidt number for
  density matrices},\ }\href {https://doi.org/10.1103/PhysRevA.61.040301}
  {\bibfield  {journal} {\bibinfo  {journal} {Phys. Rev. A}\ }\textbf {\bibinfo
  {volume} {61}},\ \bibinfo {pages} {040301} (\bibinfo {year}
  {2000})}\BibitemShut {NoStop}%
\bibitem [{\citenamefont {Cao}\ \emph {et~al.}(2023)\citenamefont {Cao},
  \citenamefont {Morelli}, \citenamefont {Rozema}, \citenamefont {Zhang},
  \citenamefont {Tavakoli},\ and\ \citenamefont {Walther}}]{cao2023genuine}%
  \BibitemOpen
  \bibfield  {author} {\bibinfo {author} {\bibfnamefont {H.}~\bibnamefont
  {Cao}}, \bibinfo {author} {\bibfnamefont {S.}~\bibnamefont {Morelli}},
  \bibinfo {author} {\bibfnamefont {L.~A.}\ \bibnamefont {Rozema}}, \bibinfo
  {author} {\bibfnamefont {C.}~\bibnamefont {Zhang}}, \bibinfo {author}
  {\bibfnamefont {A.}~\bibnamefont {Tavakoli}},\ and\ \bibinfo {author}
  {\bibfnamefont {P.}~\bibnamefont {Walther}},\ }\href@noop {} {\bibinfo
  {title} {Genuine multipartite entanglement without fully controllable
  measurements}} (\bibinfo {year} {2023}),\ \Eprint
  {https://arxiv.org/abs/2310.11946} {arXiv:2310.11946 [quant-ph]} \BibitemShut
  {NoStop}%
\bibitem [{\citenamefont {Gisin}(2007)}]{gisin2007bell}%
  \BibitemOpen
  \bibfield  {author} {\bibinfo {author} {\bibfnamefont {N.}~\bibnamefont
  {Gisin}},\ }\href@noop {} {\bibinfo {title} {Bell inequalities: many
  questions, a few answers}} (\bibinfo {year} {2007}),\ \Eprint
  {https://arxiv.org/abs/quant-ph/0702021} {arXiv:quant-ph/0702021 [quant-ph]}
  \BibitemShut {NoStop}%
\bibitem [{\citenamefont {Tavakoli}\ and\ \citenamefont
  {Gisin}(2020)}]{Tavakoli2020platonicsolids}%
  \BibitemOpen
  \bibfield  {author} {\bibinfo {author} {\bibfnamefont {A.}~\bibnamefont
  {Tavakoli}}\ and\ \bibinfo {author} {\bibfnamefont {N.}~\bibnamefont
  {Gisin}},\ }\bibfield  {title} {\bibinfo {title} {The {P}latonic solids and
  fundamental tests of quantum mechanics},\ }\href
  {https://doi.org/10.22331/q-2020-07-09-293} {\bibfield  {journal} {\bibinfo
  {journal} {{Quantum}}\ }\textbf {\bibinfo {volume} {4}},\ \bibinfo {pages}
  {293} (\bibinfo {year} {2020})}\BibitemShut {NoStop}%
\bibitem [{\citenamefont {Kocsis}\ \emph {et~al.}(2015)\citenamefont {Kocsis},
  \citenamefont {Hall}, \citenamefont {Bennet}, \citenamefont {Saunders},\ and\
  \citenamefont {Pryde}}]{Kocsis2015}%
  \BibitemOpen
  \bibfield  {author} {\bibinfo {author} {\bibfnamefont {S.}~\bibnamefont
  {Kocsis}}, \bibinfo {author} {\bibfnamefont {M.~J.~W.}\ \bibnamefont {Hall}},
  \bibinfo {author} {\bibfnamefont {A.~J.}\ \bibnamefont {Bennet}}, \bibinfo
  {author} {\bibfnamefont {D.~J.}\ \bibnamefont {Saunders}},\ and\ \bibinfo
  {author} {\bibfnamefont {G.~J.}\ \bibnamefont {Pryde}},\ }\bibfield  {title}
  {\bibinfo {title} {Experimental measurement-device-independent verification
  of quantum steering},\ }\href {https://doi.org/10.1038/ncomms6886} {\bibfield
   {journal} {\bibinfo  {journal} {Nature Communications}\ }\textbf {\bibinfo
  {volume} {6}},\ \bibinfo {pages} {5886} (\bibinfo {year} {2015})}\BibitemShut
  {NoStop}%
\bibitem [{\citenamefont {Zhao}\ \emph {et~al.}(2020)\citenamefont {Zhao},
  \citenamefont {Ku}, \citenamefont {Chen}, \citenamefont {Chen}, \citenamefont
  {Nori}, \citenamefont {Xiang}, \citenamefont {Li}, \citenamefont {Guo},\ and\
  \citenamefont {Chen}}]{Zhao2020}%
  \BibitemOpen
  \bibfield  {author} {\bibinfo {author} {\bibfnamefont {Y.-Y.}\ \bibnamefont
  {Zhao}}, \bibinfo {author} {\bibfnamefont {H.-Y.}\ \bibnamefont {Ku}},
  \bibinfo {author} {\bibfnamefont {S.-L.}\ \bibnamefont {Chen}}, \bibinfo
  {author} {\bibfnamefont {H.-B.}\ \bibnamefont {Chen}}, \bibinfo {author}
  {\bibfnamefont {F.}~\bibnamefont {Nori}}, \bibinfo {author} {\bibfnamefont
  {G.-Y.}\ \bibnamefont {Xiang}}, \bibinfo {author} {\bibfnamefont {C.-F.}\
  \bibnamefont {Li}}, \bibinfo {author} {\bibfnamefont {G.-C.}\ \bibnamefont
  {Guo}},\ and\ \bibinfo {author} {\bibfnamefont {Y.-N.}\ \bibnamefont
  {Chen}},\ }\bibfield  {title} {\bibinfo {title} {Experimental demonstration
  of measurement-device-independent measure of quantum steering},\ }\href
  {https://doi.org/10.1038/s41534-020-00307-9} {\bibfield  {journal} {\bibinfo
  {journal} {npj Quantum Information}\ }\textbf {\bibinfo {volume} {6}},\
  \bibinfo {pages} {77} (\bibinfo {year} {2020})}\BibitemShut {NoStop}%
\bibitem [{\citenamefont {Rosset}\ \emph {et~al.}(2012)\citenamefont {Rosset},
  \citenamefont {Ferretti-Sch\"obitz}, \citenamefont {Bancal}, \citenamefont
  {Gisin},\ and\ \citenamefont {Liang}}]{Rosset2012}%
  \BibitemOpen
  \bibfield  {author} {\bibinfo {author} {\bibfnamefont {D.}~\bibnamefont
  {Rosset}}, \bibinfo {author} {\bibfnamefont {R.}~\bibnamefont
  {Ferretti-Sch\"obitz}}, \bibinfo {author} {\bibfnamefont {J.-D.}\
  \bibnamefont {Bancal}}, \bibinfo {author} {\bibfnamefont {N.}~\bibnamefont
  {Gisin}},\ and\ \bibinfo {author} {\bibfnamefont {Y.-C.}\ \bibnamefont
  {Liang}},\ }\bibfield  {title} {\bibinfo {title} {Imperfect measurement
  settings: Implications for quantum state tomography and entanglement
  witnesses},\ }\href {https://doi.org/10.1103/PhysRevA.86.062325} {\bibfield
  {journal} {\bibinfo  {journal} {Phys. Rev. A}\ }\textbf {\bibinfo {volume}
  {86}},\ \bibinfo {pages} {062325} (\bibinfo {year} {2012})}\BibitemShut
  {NoStop}%
\bibitem [{\citenamefont {Sarkar}(2023)}]{Sarkar}%
  \BibitemOpen
  \bibfield  {author} {\bibinfo {author} {\bibfnamefont {S.}~\bibnamefont
  {Sarkar}},\ }\href@noop {} {\bibinfo {title} {Distrustful quantum steering}}
  (\bibinfo {year} {2023}),\ \Eprint {https://arxiv.org/abs/2306.xxxx}
  {arXiv:2306.xxxx [quant-ph]} \BibitemShut {NoStop}%
\end{thebibliography}%

\onecolumngrid

\section{Proof of Lemma}\label{AppProof}
Here, we prove Lemma 1 from the main text. 	Consider a state $\sigma$ of arbitrary dimension $D$. The state is only constrained such that it is at least $\epsilon$-close in fidelity to another, known, state $\ketbra{\psi}$. That is,
\begin{equation}\label{fiid}
\bracket{\psi}{\sigma}{\psi}\geq 1-\epsilon.
\end{equation} 
The set of states that are $\epsilon$-close to $\ket{\psi}$  is denoted $\mathcal{S}_\epsilon(\ket{\psi})$. We write the state as $\sigma=\sum_i q_i \ketbra{\phi_i}$, for some $\{q_i\}$ such that $q_i\geq 0$ and $\sum_i q_i=1$, and some pure states $\ket{\phi_i}$. Each pure state in the ensemble can be written as
\begin{equation}
\ket{\phi_i}=\sqrt{1-\xi_i}\ket{\psi}+\sqrt{\xi_i}\ket{\psi_\perp^i},
\end{equation} 
where $\ket{\psi_\perp^i}$, for each $i$, is some vector orthogonal to $\ket{\psi}$, and $0\leq \xi_i \leq 1$. Let us now define an operator  of the form
\begin{equation}\label{N}
N=(1+\mu)\ketbra{\psi}{\psi}+\eta \openone,
\end{equation}
for some parameters $\mu\geq -1$ and $\eta \geq 0$. Note that this operator is positive semi-definite by construction and that the identity is of dimension $D$. Our goal is to choose the parameters $(\mu,\eta)$ in such a way that 
\begin{align}\label{pseudopovm}
\hspace{-20mm}\forall \sigma\in \mathcal{S}_\epsilon(\ket{\psi}):  & \qquad\qquad N\succeq \sigma. 
\end{align}
We note that the average fidelity condition \eqref{fiid} now can be written as
\begin{equation}\label{fiiid}
\sum_i q_i \xi_i \leq \epsilon.
\end{equation}
We must now determine $(\mu,\eta)$ such that $R\equiv \bracket{\varphi}{(N-\sigma)}{\varphi}\geq 0$ for every $\ket{\varphi}$ and every $\sigma$ compatible with \eqref{fiiid}. Expanding $R$, we obtain
\begin{equation}
R= \eta +  |\braket{\varphi}{\psi}|^2 \left(\mu+\sum_i q_i\xi_i\right)- \sum_i q_i \xi_i |\braket{\varphi}{\psi_\perp^i}|^2- 2 \sum_i q_i\sqrt{\xi_i(1-\xi_i)} \Re\left(\braket{\varphi}{\psi}\braket{\psi_\perp^i}{\varphi}\right).
\end{equation} 
Let us now introduce variables $x=\braket{\varphi}{\psi}=|x|e^{i\theta}$	and $x_{i}=\braket{\varphi}{\psi_\perp^i}=|x_i|e^{i\theta_i}$. 	Since we require that $R\geq0$ for every choice of $(x,\{x_i\})$, we can without loss of generality choose the phases such that $\Re\left(xx_i^*\right)$ is maximal, i.e.~$\theta=\theta_i$. Without loss of generality, we can proceed with $(x,\{x_i\})$ being real parameters. Furthermore,  note that $|x|^2+|x_i|^2\leq 1$ due to conservation of probability. For any given $x$, the smallest value of $R$ is obtained by choosing $x_i$ with the same sign as $x$ and the largest possible magnitude $|x_i|$. Therefore, without loss of generality, we can restrict to $0 \leq x\leq 1$ and $x_i=\sqrt{1-x^2}$. Hence,
\begin{equation}\label{term}
R \geq \eta-\sum_i q_i\xi_i +  x^2 \left(\mu+ 2\sum_i q_i\xi_i  \right)    - 2 x \sqrt{1-x^2}\sum_i q_i\sqrt{\xi_i(1-\xi_i)}.
\end{equation}
Next, note that the function  $f(\xi)=\sqrt{\xi(1-\xi)}$ is concave on $\xi\in [0,1]$. Therefore we use Jensen's inequality, which implies $
\sum_i q_i f(\xi_i) \leq f\big(\sum_i q_i \xi_i\big)$. Defining $\xi=\sum_i q_i\xi_i$, we obtain
\begin{equation}\label{term2}
R \geq \eta-\xi  +  x^2 \left(\mu+ 2\xi  \right)  - 2 x \sqrt{1-x^2}\sqrt{\xi(1-\xi)}\equiv R'.
\end{equation}

To find the minimum, we solve $\frac{d R'}{dx}=0$. There are four possible solutions, of which only two are in the interval $x\in[0,1]$. Of these two solutions, one finds that one corresponds to the smaller value of $R$. That solution is
\begin{equation}\label{xopt}
x_\text{opt}=\frac{1}{\sqrt{2}}\sqrt{1- \frac{\mu+2\xi}{\sqrt{\mu^2+4\xi(1+\mu)}}}.
\end{equation}
Evaluating at $x=x_\text{opt}$, we have
\begin{equation}
R'=\frac{1}{2}\left(2\eta +\mu - \sqrt{\mu^2 +4\xi(1+\mu)}\right).
\end{equation}
Computing the derivative, we have
\begin{equation}
\frac{dR'}{d\xi}=-\frac{1+\mu}{\mu^2+4\xi(1+\mu)},
\end{equation}
which is non-positive. Thus, $R'$ is monotonically decreasing in $\xi$. Consequently, we must choose the largest possible value of $\xi$, which due to the fidelity condition \eqref{fiiid} simply becomes $\xi= \epsilon$. Next, we solve the equation $R'=0$ and obtain 
\begin{equation}
\eta_\text{opt}=\frac{\sqrt{\mu^2+4\epsilon(1+\mu)}-\mu}{2}.
\end{equation}	
Inserting this into Eq.~\eqref{N}, we conclude that for every $\sigma$ of arbitrary dimension $D$, such that the average fidelity is $\bracket{\psi}{\sigma}{\psi}\geq 1-\epsilon$, it holds that
\begin{equation}
\sigma \preceq (1+\mu)\ketbra{\psi}{\psi}+\frac{\sqrt{\mu^2+4\epsilon(1+\mu)}-\mu}{2} \openone_D.
\end{equation}


\section{Equal imprecision parameters}\label{AppB}
Consider the case when all imprecisions are identical, namely $\epsilon=\epsilon_{by}$.  The bound on the steering functional becomes
\begin{align}
\mathcal{B}_{\epsilon}\leq \min_{\mu\geq -1}\left(1+\mu\right)\beta_0+\frac{\chi}{2}\left(\sqrt{\mu^2+4\epsilon(1+\mu)}-\mu\right),
\end{align}
where $\chi=\sum_{x} \max_{a} \left[\sum_{b,y} c_{abxy}\right]$.  We then solve the derivative equation
\begin{equation}
\frac{d}{d\mu}\bigg[\left(1+\mu\right)\beta_0+\frac{\chi}{2}\left(\sqrt{\mu^2+4\epsilon(1+\mu)}-\mu\right)\bigg]=0.
\end{equation}
The solutions are 
\begin{align}
& \mu_{\pm}=\frac{1}{\beta_0(\chi-\beta_0)}\left(-2\beta_0\epsilon(\chi-\beta_0)\pm |\chi-2\beta_0|\sqrt{\epsilon(1-\epsilon)\beta_0(\chi-\beta_0)}\right).
\end{align}
The minimum corresponds to $\mu_-$ when $2\beta_0\geq  \chi$ and to $\mu_+$ otherwise. It turns out that in both regimes, the expression for the optimal bound becomes the same, namely 
\begin{equation}
\mathcal{B}_{\vec{\epsilon}}\leq \beta_0 -\epsilon(2\beta_0-\chi)+2\sqrt{\epsilon(1-\epsilon)\beta_0(\chi-\beta_0)}.
\end{equation}

\end{document}